\documentclass[preprint,12pt]{article}

\usepackage{amssymb}
 \usepackage{amsthm}
\usepackage{amsmath}

\usepackage{color,graphicx}
\usepackage{hyperref}

\definecolor{darkred}{rgb}{0.5,0,0.5}
\hypersetup{pdfborder={0 0 0},colorlinks=true,urlcolor=darkred,citecolor=blue,linkcolor=blue}

\begin{document}

\begin{titlepage}
\begin{center}

\hfill UG-12-35 \\ \hfill AEI-2012-055

\vskip 1cm

{\Large \bf  Supersymmetric Domain Walls}

\vskip 1cm

{\bf Eric A.~Bergshoeff\,$^1$, Axel Kleinschmidt\,$^2$  and Fabio
Riccioni\,$^3$}

\vskip 25pt

{\em $^1$ \hskip -.1truecm Centre for Theoretical Physics,
University of Groningen, \\ Nijenborgh 4, 9747 AG Groningen, The
Netherlands \vskip 5pt }

{email: {\tt E.A.Bergshoeff@rug.nl}} \\

\vskip 15pt

{\em $^2$ \hskip -.1truecm Max Planck Institute for Gravitational
Physics, Albert Einstein Institute
Am Muhlenberg 1, 14476 Potsdam, Germany\\
\&\\
International Solvay Institutes\\
Campus Plaine C.P. 231, Boulevard du Triomphe, 1050 Bruxelles,
Belgium \vskip 5pt }

{email: {\tt axel.kleinschmidt@aei.mpg.de }} \\

\vskip 15pt

{\em $^3$ \hskip -.1truecm
 INFN Sezione di Roma,  Dipartimento di Fisica,\\ Universit\`a di Roma ``La Sapienza'',\\ Piazzale Aldo Moro 2, 00185 Roma, Italy
 \vskip 5pt }

{email: {\tt Fabio.Riccioni@roma1.infn.it}} \\

\end{center}

\vskip 0.5cm

\begin{center} {\bf ABSTRACT}\\[3ex]
\end{center}

We classify the half-supersymmetric ``domain walls'', i.e.~branes of
  codi\-men\-sion one, in toroidally compactified IIA/IIB string theory
  and show to which gauged supergravity theory each of these domain walls belong.
  We use as input the requirement of supersymmetric Wess-Zumino terms, the  pro\-per\-ties of the
  $E_{11}$ Kac-Moody algebra and the embedding tensor formalism. We show that the number
  of half-supersymmetric domain walls is  a multiple of the number of corresponding central charges in the supersymmetry
  algebra,  where the multiplicity is related to  the degeneracy of the BPS conditions.

\end{titlepage}

\newpage
\setcounter{page}{1} \tableofcontents

\newpage

\setcounter{page}{1} \numberwithin{equation}{section}


\section{Introduction}
\label{introduction}

Domain walls are branes of codimension one, i.e.~they have a single
transverse direction. They play an important role in a wide range of
situations such as  setting up brane-world
scenarios~\cite{Kaloper:1999sm} and describing the renormalization
group flow in an AdS/CFT setting~\cite{Skenderis:1999mm}. A
distinguishing feature of domain walls is that their existence
within a supergravity context,  unlike that of other branes,
requires the use of a {\sl deformed} supergravity theory where the
deformation parameter can be a mass parameter (``massive''
supergravity) or a gauge coupling constant (``gauged''
supergravity). These deformed supergravity theories generically
contain a potential for the scalar fields which is needed to realize
the domain-wall solutions. The same potential is also needed to
allow interesting cosmological solutions. The study of these
cosmological solutions is relevant for our efforts to extract an
expanding de Sitter solution out of a string theory
compactification. Domain walls and cosmologies are related to each
other via the ``domain-wall/cosmology'' correspondence
\cite{Skenderis:2007sm}. In view of the above remarks it is
important to classify all supersymmetric domain walls and determine
which deformed supergravity theory they are related to.

The prime example of a half-supersymmetric domain wall is the
D8-brane of IIA string theory. This brane is electrically charged
with respect to the Ramond-Ramond (RR) 9-form potential $C_9$, thus
leading to a solution such that the corresponding 10-form field
strength $G_{10}$ is non-vanishing and proportional to a mass
parameter $m$ (that is constant by virtue of the Bianchi identity).
This means that the presence of the D8-brane source induces a
cosmological constant in the theory, and this corresponds to a
domain-wall solution of the Romans IIA theory \cite{Romans:1985tz},
whose explicit form, in Einstein frame, is \cite{Bergshoeff:1996ui}
\begin{eqnarray}
\label{D8met}
ds^2 &=& H^{9/8}dy^2 + H^{1/8} dx^\mu dx^\nu \eta_{\mu\nu} \,,\nonumber \\ [.2truecm]
e^{\phi} &=& H^{5/4}\,, \label{D8solutionromans}\\ [.2truecm]
C_{01\cdots 8} &=& \pm H^{-1}\,,\hskip .5truecm {\rm or} \hskip .5truecm  m = \pm H^\prime\,. \nonumber
\end{eqnarray}
Here $y$ indicates the transverse direction of the domain wall, the
prime indicates a differentiation with respect to $y$ and $H(y)$ is
a harmonic function of $y$. The Romans deformation is a massive
deformation and not a gauge deformation,\,\footnote{The
$\mathbb{R}^+$-scaling symmetry of the theory cannot be gauged since
the RR 1--form $C_1$ has a non-zero weight under this symmetry.}
which  stems from the RR 1-form $C_1$  transforming with a shift,
proportional to $m$, under the gauge parameter $\Sigma_1$ of the
Neveu-Schwarz/Neveu-Schwarz (NS--NS) 2-form $B_2$. Therefore, $C_1$
is ``eaten up'' by $B_2$ and the two potentials $(C_1,B_2)$ together
form a so-called St\"uckelberg pair describing a massive 2-form. A
similar St\"uckelberg mechanism happens for the dual potentials
$(D_6,C_7)$ where $D_6$ is eaten up by $C_7$. The RR 3-forms $C_3$
and its dual 5-form $C_5$ remain massless. The situation is
summarized in Table \ref{table:Romans}.\footnote{ We have not
indicated the RR 9-form $C_9$ in Table \ref{table:Romans} since it
does not describe a physical degree of freedom. In fact, the dual of
its curvature is proportional  to the mass parameter $m$. We have
neither indicated the two 10-form potentials that can be added to
the IIA supergravity multiplet \cite{Bergshoeff:2010mv}. They do not
couple to half-supersymmetric space-filling branes.} The deformation also induces
a change in the Hodge duality relations for the massive forms. They
take on the form of massive duality relations that are roughly of
the form $dB_2 = m*C_7$. That is, the rank of  dual forms is shifted
by one for massive fields as compared to massless duality.

\begin{table}[h]
\begin{center}
{\small
\begin{tabular}{|c||c|c|c|c|c|c|}
\hline\rule[-1mm]{0mm}{6mm}
gauging &$C_1$ &$B_2$&$C_3 $&$C_5$&$D_6$&$C_7$\\[.1truecm]
\hline \rule[-1mm]{0mm}{6mm}  $m$ &gauged&massive &massless&massless&gauged&massive\\[.05truecm]
\hline
\end{tabular}
}
\end{center}
\caption{\sl The  Romans deformation in ten dimensions corresponds
to a minimal gauging, leading to the elementary D8--brane domain-wall solution of IIA supergravity. ``Gauged'' means ``eaten up'' by
the
 neighbouring form to the right.}\label{table:Romans}
\end{table}
\bigskip

The existence of the D8-brane in IIA string theory can be
anticipated from the potentials of the massless IIA supergravity and
their gauge transformations. At  leading order the D8-brane couples
to the pull-back of the RR 9-form potential $C_9$ via a Wess-Zumino
(WZ) term. This term by itself is not gauge-invariant because $C_9$
not only transforms into a total derivative under its own gauge
transformation but it also transforms to the curvature $H_3=dB_2$ of
$B_2$:
\begin{equation}\label{general}
\delta C_9 = d\lambda_8 + H_3\lambda_6\,.
\end{equation}
Moreover,  the only worldvolume field introduced so far is the
single embedding scalar corresponding to the transverse direction of
the domain wall. This scalar by itself does not fill a
supermultiplet on the nine-dimensional worldvolume of the D8-brane.
Both problems, the gauge-invariance of the WZ term and the
supersymmetry on the worldvolume, can be solved simultaneously by
introducing a worldvolume gauge vector $b_1$ transforming with
respect to the pull-back of the gauge parameter of the 2-form as
$\delta b_1 = -\Sigma_1$,  which implies that ${\cal F}_2 \equiv
db_1 + B_2$ is gauge-invariant. A gauge-invariant WZ term is then
given by\,\footnote{The fact that ${\cal F}_2$ occurs
non-polynomially in the WZ term  is related to the fact that ${\cal
F}_2$, like $B_2$, has scaling weight zero.}
\begin{equation}
{\cal L}_{\rm WZ} = e^{{\cal F}_2}\, C\,. \label{DbraneWZtermD=10}
\end{equation}
Here we use the standard notation where all RR potentials are
contained in the formal sum $C\equiv C_1 + C_3 + C_5 + \dots$. To
obtain the WZ term for the D8-brane one should project
eq.~\eqref{DbraneWZtermD=10} onto 9-forms. It is easily seen that
this WZ term is invariant under the transformations $\delta C=
d\lambda + H_3 \lambda$ which generalizes eq.~\eqref{general} to the
other RR potentials, and using the Bianchi identity $d{\cal F}_2 =
H_3$. At the same time the introduction of a vector $b_1$ on the
worldvolume, together with the transverse embedding scalar, fills a
nine-dimensional vector multiplet. We stress that when constructing
this gauge-invariant WZ term for the D8--brane one considers the
transformation rules of the {\sl un-deformed} IIA supergravity,
which implies  that the curvature $G_{10}$ of the RR 9-form $C_9$ is
zero. This means that one is only considering here the D8-brane as a
test brane thereby ignoring the back-reaction of the D8-brane on the
supergravity background,\,\footnote{One can also consider the
D-branes as  test branes in a {\sl deformed} supergravity
background, see, e.g., \cite{Bergshoeff:1997cf}. } which would
indeed lead to the solution \eqref{D8solutionromans} of the Romans
theory.

\begin{table}[t]
\begin{center}
\hskip -1.3truecm
\begin{tabular}{|c|c|c|}
\hline \rule[-1mm]{0mm}{6mm}
$D$&U&T\\[.1truecm]\hline
\hline \rule[-1mm]{0mm}{6mm} IIA&$\mathbb{R}^+$ &$1$ \\[.05truecm]
\hline \rule[-1mm]{0mm}{6mm} IIB& $SL(2, \mathbb{R})$&$1$ \\[.05truecm]
\hline \rule[-1mm]{0mm}{6mm} 9&$SL(2,\mathbb{R})\times \mathbb{R}^+$&$SO(1,1)$ \\[.05truecm]
\hline \rule[-1mm]{0mm}{6mm} 8&$SL(3,\mathbb{R})\times SL(2,\mathbb{R})$&$SL(2,\mathbb{R})\times SL(2,\mathbb{R})$\\[.05truecm]
\hline \rule[-1mm]{0mm}{6mm} 7&$SL(5,\mathbb{R})$&$SL(4,\mathbb{R})$\\[.05truecm]
\hline \rule[-1mm]{0mm}{6mm} 6&$SO(5,5)$&$SO(4,4)$\\[.05truecm]
\hline \rule[-1mm]{0mm}{6mm} 5&${E}_{6(6)}$&$SO(5,5)$ \\[.05truecm]
\hline \rule[-1mm]{0mm}{6mm} 4&${E}_{7(7)}$&$SO(6,6)$ \\[.05truecm]
\hline \rule[-1mm]{0mm}{6mm} 3&${E}_{8(8)}$&$SO(7,7)$\\[.05truecm]
\hline
\end{tabular}
\end{center}
 \caption{\sl This table indicates the continuous global U-duality and T-duality symmetries in  dimensions $3\le D\le 10$.
\label{table:UTsymmetries}}
\end{table}

The existence of a gauge-invariant WZ term is a necessary but not
sufficient requirement for the existence of the D8-brane. The full
worldvolume action  also contains kinetic terms for the worldvolume
fields. This worldvolume action describes the dynamics of a single
domain wall which does not constitute a finite-energy object by
itself. For that one needs to introduce more domain walls and
orientifolds as well. However, without going into the details of the
precise construction, the requirement of a gauge-invariant WZ term
consistent with supersymmetry on the worldvolume is a useful
criterion which was applied in
\cite{Bergshoeff:2010xc,Bergshoeff:2011zk,Bergshoeff:2011qk,Bergshoeff:2012ex}
to determine all  the supersymmetric branes that occur in IIA/IIB
string theory compactified on a torus. In particular, this analysis
shows that for the case of branes of codimension 2, 1 or 0 (that is,
defect branes, domain walls and space-filling branes) the number of
supersymmetric branes is lower than the dimension of the U-duality
representation of the corresponding form fields. The same conclusion
was reached by the analysis of \cite{Kleinschmidt:2011vu}, where the
supersymmetric branes were counted as those corresponding to
potentials associated to the real roots of the Kac-Moody algebra
$E_{11}$   \cite{West:2001as}. As far as domain walls are concerned,
the outcome of this analysis, given in each dimension in terms of
representations of the U-duality and T-duality groups (see Table
\ref{table:UTsymmetries})  is summarized in Table \ref{table1}
\cite{Kleinschmidt:2011vu,Bergshoeff:2012ex}. The U-duality
representations of the $p$-form fields in dimensions $D\geq 3$ can
be obtained either by using the tensor hierarchy
formalism~\cite{deWit:2005hv,deWit:2008ta}, the Kac-Moody algebra
$E_{11}$~\cite{Riccioni:2007au,Bergshoeff:2007qi}, a Borcherds
algebra approach~\cite{HenryLabordere:2002dk,Palmkvist:2011vz} or
superspace methods~\cite{Greitz:2011vh}.

\begin{table}[t]
\hskip -1.5truecm
{\small
\begin{tabular}{|c|c||c|c|c|c|c|c|}
\hline\rule[-1mm]{0mm}{6mm}
$D$&U repr.&$\alpha=0$&$\alpha=-1$&$\alpha=-2$&$\alpha=-3$&$\alpha=-4$&$\alpha=-5$\\[.1truecm]
\hline \rule[-1mm]{0mm}{6mm}IIA&1&&$1$&&&&\\[.05truecm]
\hline \rule[-1mm]{0mm}{6mm}9&$2\subset {\bf 3}$&&$1$&$-$&$1$&&\\[.05truecm]
\hline \rule[-1mm]{0mm}{6mm}8&$6\subset {\bf (6,2)}$&&$ {\bf (1,2)}$&$-$&$4\subset {\bf (3,2)}$&&\\[.05truecm]
\hline \rule[-1mm]{0mm}{6mm}7&$20\subset {\bf {\overline {40}}}$&&{\bf 4}&$ 4\subset {\bf 10}$&$12\subset \overline{{\bf 20}}$&&\\[.05truecm]
\cline{2-8} \rule[-1mm]{0mm}{6mm}&$5\subset \overline{{\bf 15}}$&&&$4\subset \overline{{\bf 10}}$&$-$&{\bf 1}&\\[.05truecm]
\hline \rule[-1mm]{0mm}{6mm}6&$80\subset {\bf 144}$&&${\bf 8}_{\rm S}$&$32\subset {\bf 56}_{\rm C}$&$32\subset {\bf 56}_{\rm S}$&${\bf 8}_{\rm C}$&\\[.05truecm]
\hline \rule[-1mm]{0mm}{6mm}5&$216\subset {\bf 351}$&&$\overline{{\bf 16}}$&$80\subset {\bf 120}$&$80\subset  {\bf 144}$&$40\subset {\bf 45}$&\\[.05truecm]
\hline \rule[-1mm]{0mm}{6mm}4&$576\subset {\bf 912}$&&${\bf 32}$&$160\subset {\bf 220}$&$192\subset {\bf 352}$&$160\subset{\bf 220}$&${\bf 32}$\\[.05truecm]
\hline \rule[-1mm]{0mm}{6mm}3&$2160\subset {\bf 3875}$&${\bf 1}$&$\overline{{\bf 64}}$&$280\subset {\bf 364}$&$448\subset {\bf 832}$&$560\subset {\bf 1001}$&$448\subset \overline{{\bf 832}}$\\[.05truecm]
\rule[-1mm]{0mm}{6mm} &&&&&&$14\subset {\bf 104}$&\\[.05truecm]
\rule[-1mm]{0mm}{6mm} $\alpha \le -6$&&&&&$280\subset {\bf 364}_{-6}$&${\bf 64}_{-7}$&${\bf 1}_{-8}$\\[.05truecm]
\hline
\end{tabular}
}
\caption{\sl The number of supersymmetric domain walls  in different
dimensions. The 7D case is discussed in Subsection 2.1.2. The
representations at the right of the double vertical line indicate
T-duality representations. The number $\alpha$ denotes the scaling
of the mass $M$ of the brane with the string coupling $g_s$, i.e.,
$M \sim g_{\text s}^{-\alpha}$. In the last row, where the domain
walls with $\alpha\le -6$ occur,
 we have indicated  the value of $\alpha$ with  a sub-index.}\label{table1}
\end{table}

This analysis, which gives the half-supersymmetric domain walls that
can be introduced as probes in the undeformed supergravity theories
in any dimensions, leaves unsolved the problem of determining which
deformation of the supergravity theories these branes  induce when
the back-reaction with supergravity is taken into account. In the
ten-dimensional case it was not difficult to guess which deformed
IIA supergravity theory the D8 brane is a solution of, as  there is
only one deformation  characterized by the single parameter $m$. In
$D<10$ dimensions,  all  deformations of the maximal supergravity
theories turn out to be gauged supergravities, and they are nicely
classified in a U-duality covariant way by the so-called embedding
tensor formalism \cite{deWit:2005hv}. The embedding tensor is an
object belonging to a given representation of the U-duality group
describing how the gauge group is embedded inside U-duality.
Correspondingly, the theory admits $(D-1)$-form potentials belonging
to representations which are dual to those of the embedding tensor,
and whose $D$-form field strengths are related to the embedding
tensor by duality
\cite{Riccioni:2007au,Bergshoeff:2007qi,deWit:2008ta}. However,
since there are more $(D-1)$-form potentials than there are
half-supersymmetric domain walls that allow supersymmetric WZ terms,  not all gauged supergravities
admit such half-supersymmetric domain-wall solutions. One of the purposes
of this work is to find out which gauged supergravities correspond
to  half-supersymmetric domain walls allowing supersymmetric WZ terms.

A partial classification of half-supersymmetric domain-wall
solutions of maximal supergravity theories was performed in
\cite{Cowdall:1996tw} by considering all possible gauged theories in
$D$ dimensions that arise as Scherk-Schwarz reductions from $D+1$
dimensions. Indeed, by considering a ``vertical'' dimensional
reduction of a defect-brane solution which is magnetically charged
under a given axion in $D+1$ dimensions, one obtains a domain-wall
solution, but at the same time one has to impose for consistency
that the axion depends linearly on the compactification coordinate,
thus leading to a gauged theory in $D$ dimensions. All gaugings in
$D \geq 7$ obtained with this method, and the corresponding domain
wall solutions, were classified. We will comment at
 several places in the paper when our general analysis reproduces the results of \cite{Cowdall:1996tw}.

There is one subtlety in $D<10$ that does not occur in $D=10$
dimensions: it turns out that there are  gauged  supergravities that
allow for half-supersymmetric domain-wall solutions that do not
correspond to the half-supersymmetric branes  following from the WZ
term analysis. To distinguish between the different domain walls we
will call the ones that do satisfy the WZ term criterion the {\sl
elementary}  domain walls. The dynamics of the other domain walls
cannot be described by a  supersymmetric worldvolume action, and
they are interpreted as bound states of the elementary domain walls.
We have already seen in the case of the Romans theory that deforming
a massless supergravity theory leads to a rearrangement of the
degrees of freedom (see Table \ref{table:Romans}). In this work we
will define the {\sl minimal} gaugings as those for which this
rearrangement is minimal, i.e., the minimal number of fields changes
behaviour. Correspondingly the subgroup of  the U-duality group that
remains as a global symmetry of the deformed theory is maximal. We
will see that all the gaugings that allow for elementary domain-wall
solutions are minimal. We will also determine the gaugings that
allow for  domain-wall solutions  describing threshold bound states,
that is bound states preserving the same amount of supersymmetry as
the elementary domain walls, as well as the gaugings that allow for
domain-wall solutions describing non-threshold bound states, that
have less supersymmetry.

The existence of threshold bound states   shows that there are
different elementary domain walls that satisfy the same BPS
condition. This degeneracy does not occur for branes with three or
more transverse directions. In general, the possible BPS conditions
are in 1-1 correspondence with the central charges of the
supersymmetry algebra. Therefore, for branes with three or more
transverse directions one finds a 1-1 correspondence between these
branes and the central charges \cite{de Azcarraga:1989gm}. This is
not the case for branes with less than three transverse directions.
For defect branes, i.e.~branes with two transverse directions, one
finds a double degeneracy: each BPS condition is satisfied by two
defect branes which are related to each other by an  S-duality
transformation \cite{Bergshoeff:2011se}. In this work we will spell
out what  the precise degeneracy structure is in the case of domain
walls.

It turns out that several of the results derived in this work can
be understood from the point of view of the $E_{11}$ Kac-Moody
algebra \cite{West:2001as}. This applies in particular to   the
classification of the half-supersymmetric domain walls,  the
analysis of the domain-wall solutions and the structure of the BPS
conditions. We will at several places in the paper present this
alternative $E_{11}$ point of view.

This work is organized as follows. In section 2 we classify the
elementary supersymmetric domain walls by requiring the existence of
a gauge-invariant WZ term that is consistent with supersymmetry on
the worldvolume. We discuss an alternative derivation making use of
the real roots of the $E_{11}$ Kac-Moody algebra. In section 3 we
discuss which gauged supergravities correspond to these elementary
supersymmetric domain walls. To illustrate our methods we present
explicit results for the 9D, 8D and 7D gaugings. Next, in section 4,
we discuss the supersymmetric domain-wall solutions to these gauged
supergravities. We point out that there exists a wider class of
gauged supergravities that allow many more supersymmetric domain
wall solutions. These solutions correspond to threshold and
non-threshold bound states of the elementary domain walls discussed
in the previous two sections. We will discuss the same solutions
from an $E_{11}$ point of view. In section 5 we show that there is a
relation between the number of elementary domain walls and the
number of 2-form ($4 \le D \le 10$) and 1-form (3D) central charges
in the supersymmetry algebra after one takes into account the
degeneracy of the BPS conditions involved. We discuss the relation
between these BPS conditions and $E_{11}$.  Finally, in section 6 we
present our conclusions.

\section{Classifying Supersymmetric Domain Walls}

In this section, we set the scene by reviewing how to classify
supersymmetric domain walls by two different routes. The first is
the analysis of supersymmetric WZ terms as a necessary condition for
the existence of such domain
walls~\cite{Bergshoeff:2010xc,Bergshoeff:2011zk,Bergshoeff:2011qk,Bergshoeff:2012ex}.
The second approach uses properties of the corresponding roots of
the Kac-Moody algebra $E_{11}$~\cite{Kleinschmidt:2011vu}. We phrase
our discussion in a U-duality covariant way.

The form fields of the maximal supergravity theories in any
dimension $D$, which include the propagating forms $A_p$, with $p
\leq [ D/2 ] -1$, their magnetic duals $A_{D-p-2}$ together with the
non-propagating forms $A_{D-1}$ and $A_D$,\,\footnote{Although
non-propagating, these fields can be introduced in the undeformed
supersymmetry algebra~\cite{Bergshoeff:2010mv,Bergshoeff:2005ac}.}
were classified in
\cite{Riccioni:2007au,Bergshoeff:2007qi,deWit:2008ta} in terms of
their U-duality representations.  The WZ-term analysis of
\cite{Bergshoeff:2010xc,Bergshoeff:2011zk,Bergshoeff:2011qk,Bergshoeff:2012ex},
and the $E_{11}$  analysis of  \cite{Kleinschmidt:2011vu} led to the
conclusion that while for forms of rank less than $D-2$, that is for
branes of codimension greater than 2, there are as many branes as
the dimension of the U-duality representations of the corresponding
fields, for branes of codimension 2, 1 and 0 the following results
hold:

\begin{itemize}
\item When the representation is reducible, not all the corresponding irreducible
representations are associated to branes. The branes always correspond to the
highest-dimensional irreducible representation of the associated form, with
the exception of the 5-branes in $D=6$, $D=7$ and $D=8$, in which case the vector
branes (that is the branes supporting a worldvolume vector multiplet) belong to
the highest dimensional irreducible representation while the tensor branes
(that is the branes supporting a worldvolume tensor multiplet) belong to the second highest-dimensional one.

\item For a given irreducible representation, there are fewer supersymmetric
branes than components of the representation.
More precisely, the half-supersymmetric branes belong in all cases to the
highest-weight orbit of the corresponding representation  \cite{Kleinschmidt:2011vu,Bergshoeff:2012ex},
but while the number of branes of codimension greater than 2 are as many as the dimension of the
representation, for branes of codimension 2,1 and 0 the constraints that define the highest-weight orbit
are stronger and one always gets fewer branes than the number of component of the corresponding representation.

\end{itemize}

In the remainder of this section, we will first review how to derive
the number of half-supersymmetric branes from the analysis of the WZ
term~\cite{Bergshoeff:2010xc,Bergshoeff:2011zk,Bergshoeff:2011qk,Bergshoeff:2012ex}.
This will be done by deriving as an example the number of defect
branes and domain walls in seven dimensions. We will then review how
the same result can be obtained by counting the real roots of the
$E_{11}$ algebra \cite{Kleinschmidt:2011vu}. In Table \ref{table1}
we have listed the number of domain walls resulting from this
analysis. For completeness, we give the decomposition under T-duality
of the U-duality representations and of the corresponding number of
branes.

\subsection{Domain walls and supersymmetric WZ terms}

We consider maximal supergravity theories in $D\geq 3$ space-time
dimensions, whose global symmetries  $G=E_{11-D}$ are listed in the
second column of  Table~\ref{table:UTsymmetries}. In order to write
down a supersymmetric and gauge invariant WZ term for a $p$-brane of
the $D$-dimensional theory one has to consider at leading order the
pull-back to the worldvolume  of the appropriate $(p+1)$-form
potential. Given that such field transforms with respect to the
gauge parameters of the lower-rank fields, this term alone cannot be
gauge-invariant, and one has to add terms of the form $A \wedge
{\cal F}$, where ${\cal F}$ are the field-strengths of suitably
introduced worldvolume fields. The construction was reviewed in the
Introduction for the case of D-branes in ten dimensions, where the
gauge transformations of the RR fields force the introduction of a
worldvolume vector, so that the resulting WZ term
\eqref{DbraneWZtermD=10} is gauge-invariant. The necessary condition
for the $p$-brane to be supersymmetric is then that the worldvolume
fields (including the transverse scalars) that couple to this brane
fill out the bosonic sector of a supermultiplet of the corresponding
world-volume supersymmetry, that is either a vector (any $p$) or a
tensor ($p=5$) multiplet. In the case of the D-branes in 10
dimensions this is indeed the case as one always gets a vector plus
$10-p-1$ transverse scalars, which is indeed a vector multiplet in
$p+1$ dimensions.

We will now review how this works explicitly in $D=7$. In this case,
the global symmetry group is $SL(5,\mathbb{R})$ and the form fields are
($M,N,P=1,\ldots,5$)
\begin{center}
\begin{tabular}{cl}
$A_{1\,[MN]}$ & 1-form fields in the ${\bf \overline{10}}$\\ [.1truecm]
$A_2^M$ & 2-form fields in the ${\bf 5}$\\ [.1truecm]
$A_{3\,M}$ & 3-form fields in the ${\bf \bar{5}}$\\ [.1truecm]
$A_4^{[MN]}$ & 4-form fields in the ${\bf 10}$\\ [.1truecm]
$A_{5\,M}{}^N$ & 5-form fields in the (adjoint) ${\bf 24}$\\ [.1truecm]
$A_{6\,(MN)}$, $A_6{}^{[MN],P}$ & 6-form fields in the ${\bf \overline{15}}\oplus {\bf \overline{40}}$
\end{tabular}
\end{center}
The  6-form in the ${\bf \overline{40}}$ satisfies the irreducibility constraint $A_6^{[MN,P]}=0$.
There are also 7-form fields in the ${\bf 5}\oplus {\bf 45}\oplus {\bf 70}$, but they will be of no importance
for our discussion. We will only write down the leading WZ terms, that is the terms of the form $A + A\wedge {\cal F}$,
and not terms which are higher order in ${\cal F}$ (although we know that such higher order terms are needed for gauge
invariance). Besides, we will not determine the actual coefficient of each term: we will assume that if a given term
can be written, it will actually occur in the WZ term with non-zero coefficient.
We will now proceed with the analysis of the WZ terms of the defect branes and the domain walls in $D=7$.

\subsubsection{Example: $D=7$ defect branes}

We start by describing the WZ term analysis for $4$-branes (defect
branes), which are charged with respect to the 5-form field
$A_{5M}{}^N$. The WZ term is  of the form
\begin{align}
\label{D=7defect}
\mathcal{L}_{WZ}^{p=4} \sim A_{5\,M}{}^N +A_4{}^{NP} \mathcal{F}_{1\,PM} + A_{3\,M}\mathcal{F}_{2}{}^N -\frac15\delta_M^N \left(A_4{}^{QP} \mathcal{F}_{1\,PQ} + A_{3\,P}\mathcal{F}_{2}{}^P\right).
\end{align}
We have not written out terms containing ${\cal F}_3$ and ${\cal
F}_4$,  because we assume that they are related to ${\cal F}_2$ and
${\cal F}_1$  by five-dimensional world-volume Hodge
duality.\,\footnote{We assume that the field-strengths ${\cal
F}_{n}$ and ${\cal F}_{p+1-n}$ are related by worldvolume Hodge
duality.  We will make a similar assumption  for the other WZ terms
discussed in this section.} The field strengths $\mathcal{F}_p$ are
the field strengths of world-volume $(p-1)$-form fields augmented by
$p$-form St\"uckelberg shifts of the pull-backs. More precisely
\begin{align}
\mathcal{F}_{1\,PQ} =d a_{0\,PQ} + A_{1\,PQ}
\end{align}
with world-volume scalars $a_{0\,PQ}$ in the ${\bf \overline{10}}$
of $SL(5,\mathbb{R})$. Similarly, ${\cal F}_2^M$ is a gauge-invariant
field-strength for the worldvolume field $a_1^M$.
 The last parenthesis in (\ref{D=7defect}) is needed to ensure that the WZ term is in the traceless adjoint of $SL(5,\mathbb{R})$.

In order to determine which components of (the pull-back of)
$A_{5\,M}{}^N$ couple to supersymmetric branes, we will now consider
$A_{5\,M}{}^N$ for fixed $M$ and $N$. By analyzing eq.~\eqref{D=7defect}, one can see that in order to describe a single
vector multiplet on the worldvolume (that is one vector and five
scalars), one has to impose that $M \neq N$, so that the term in
parenthesis is not present. Then, the term $A_4^{NP} {\cal F}_{1
PM}$ gives a scalar for each allowed $P$, which is three
possibilities because $P$ is different from both $M$ and $N$. The
term $A_{3M } {\cal F}_2{}^N$ gives one vector. Finally, there are
two transverse scalars, making up a total of one vector and five scalars.
This  is the right field content for a vector supermultiplet on the
five-dimensional world-volume and the necessary criterion for a
supersymmetric $4$-brane is fulfilled.

On the other hand, if $M=N$, the last parenthesis in
(\ref{D=7defect}) does not vanish and there are many more fields
that couple to the world-volume theory. In fact, all $5$ vector
fields contribute as do all the $10$ scalars; these cannot be
grouped into world-volume supermultiplets and therefore the WZ term
cannot be supersymmetrized. The case $M=N$ does not correspond to a
supersymmetric brane. In all, only $20$  out of the $24$ components
of $A_{5\,M}{}^N$ couple to supersymmetric branes. These components
fill up the highest-weight orbit in the ${\bf 24}$.

\subsubsection{Example: $D=7$ domain walls}

We now proceed to the case of interest here, namely domain walls in
$D=7$. There are two distinct cases to consider since the $6$-forms
come in two different representations. The WZ term on the
six-dimensional world-volume for the ${\bf \overline{15}}$
representation is
\begin{align}
\label{D=7dw15}
\mathcal{L}_{WZ}^{p=5,{\bf \overline{15}}} \sim A_{6\,(MN)}+A_{5\,(M}{}^P\mathcal{F}_{1\,N)P} + A_{3\,(M}\mathcal{F}_{3\,N)},
\end{align}
where $\mathcal{F}_{3\,N}$  is the field-strength of a new world-volume 2-form field $a_{2\,N}$.  Following our worldvolume duality assumption, we must assume that this field-strength  enjoys world-volume Hodge self-duality. Now the counting of world-volume fields works as follows. If $M=N$, there is a single self-dual tensor field from the last term, while the summation index $P$ has to be different from $M=N$ and there are therefore four scalar fields from the middle term. Together with the single transverse scalar this gives a self-dual tensor plus five scalars which is exactly the right content for a tensor multiplet on the six-dimensional world-volume. There are five choices for $M$ (equal to $N$) so that there are five supersymmetric tensor domain walls, again related to the highest-weight orbit.

If $M\neq N$, one obtains two tensor fields and seven scalar fields
from (\ref{D=7dw15}). Together with the transverse scalar these do
not form supermultiplets on the world-volume and therefore there is
no supersymmetric domain wall in this case. In summary, only $5$ out
of the 15 6-forms couple to (elementary) supersymmetric tensor
domain walls.

Turning to the 6-forms in the ${\bf \overline{40}}$ of $SL(5,\mathbb{R})$, the WZ term looks like
\begin{align}
\label{D=7dw40}
\mathcal{L}_{WZ}^{p=5,{\bf \overline{40}}} &\sim A_{6}^{MN,P}+A_{5\,Q}{}^P\mathcal{F}_{1\,RS}\epsilon^{MNQRS} + A_{4}^{MN}\mathcal{F}_{2}^{P}\nonumber\\
&\quad- \left(A_{5\,Q}{}^{[P}\mathcal{F}_{1\,RS}\epsilon^{MN]QRS} + A_{4}^{[MN}\mathcal{F}_{2}^{P]}\right),
\end{align}
where the second line is needed to ensure that the irreducibility constraint of the ${\bf\overline{40}}$ is satisfied. When counting the world-volume fields one has to distinguish between the case $P=M$ (or equivalently $P=N$) and the case where all three indices are different. Starting with the former case when $P=M$, the second line in (\ref{D=7dw40}) vanishes and the last term of the first line shows that there is a single vector field. Moreover, the antisymmetric summation indices $R$ and $S$ can take only three different values, so that there are three scalar fields.  Together with the transverse scalar this gives four scalar fields. This is precisely the right number for a half-maximal vector multiplet in six world-volume dimensions. Counting the number of supersymmetric domain walls thus obtained, we find $20$, ten from when $P=M$ and ten from when $P=N$ in the antisymmetric pair $[MN]$.

Performing the analysis in the case when all indices $M,N,P$ are different, one ends up with a field content that does not fit into supermultiplets. In total there are then $20$ supersymmetric domain walls in the ${\bf \overline{40}}$, corresponding to the dimension of the highest-weight orbit.

A similar analysis can be performed in all dimensions. The result is
given in  Table~\ref{table1}. This table lists the U-duality and
T-duality representations of all elementary supersymmetric domain
walls that possess a supersymmetrizable WZ term. Note that only for
$D=3,4,6$, where we have real representations,  an elementary
supersymmetric domain wall with given $\alpha$ transforms under
S-duality  into another domain wall with another value
$\alpha^\prime$ given by:
\begin{equation}
\alpha^\prime = - \alpha - 4\,\frac{D-1}{D-2}\,.
\end{equation}
For the other dimensions S-duality does not commute with T-duality and the transformations properties are more complicated.

\subsection{Supersymmetric domain walls and $E_{11}$}

The same classification of supersymmetric domain walls can be
obtained independently from an analysis of the $E_{11}$ roots
associated with the space-time $p$-forms~\cite{Kleinschmidt:2011vu}.
The infinite-dimensional Lorentzian Kac-Moody algebra $E_{11}$
reproduces nicely the tensor hierarchy of $p$-form fields that
occurs in maximal
supergravity~\cite{Riccioni:2007au,Bergshoeff:2007qi}. In this
language, one can obtain all the $p$-form fields in $D$ space-time
dimensions and in a given representation of the U-duality group
$E_{11-D}$ by decomposing the adjoint representation of $E_{11}$
under its $E_{11-D}\times GL(D,\mathbb{R})$ subalgebra. This
decomposition produces an infinite number of fields but only a
finite number of $p$-forms along with their U-duality
representation, which are indeed the $p$-forms of the
$D$-dimensional maximal supergravity theory.

Together with the above decomposition of the adjoint one also
obtains root vectors $\alpha$ of the $E_{11}$ algebra that are
associated with the various components of the $p$-forms. In order to
decide which of these correspond to supersymmetric branes one has to
recall that the inner product on the space of root vectors of
$E_{11}$ is Lorentzian (whence the name Lorentzian Kac-Moody
algebra). This means that root vectors can be either space-like (and
are then called {\em real roots}), light-like (and are then called
{\em null roots}) or time-like (and are then called {\em purely
imaginary}). Often the last two cases are combined such that one is
left with only the distinction between real ($\alpha^2>0$) and
imaginary roots ($\alpha^2\leq 0$).

A given $p$-form transforms under U-duality such that in general the
root vectors of the components can be either real or imaginary. The
simple rule for classifying supersymmetric solutions is now that
only those components that are associated with {\it real} roots
correspond to supersymmetric branes whereas those components that
are associated with imaginary roots are not supersymmetric. The
solutions corresponding to real roots were discussed from a coset
model point of view
in~\cite{Englert:2003py,Englert:2004it,West:2004st,Cook:2004er}.
That the solutions for imaginary roots are not supersymmetric was
explicitly checked in a particular representative case
in~\cite{Houart:2011sk}.

The real roots relevant for domain walls can be easily generated and
classified by using the language of orbits under the U-duality
group. They are always in the orbit of the highest weight of a given
U-duality representation of the $(D-1)$-forms (if the highest weight
is a real root). This can be viewed alternatively as the Weyl group
orbit of the highest weight, similar to the analysis
in~\cite{Obers:1998fb}. For example, in the $D=7$ case that was
discussed above, the highest weight of the ${\bf \overline{15}}$
six-forms $A_{MN}$ is given by $A_{1\,1}$ (by choosing an ordering
of the five directions of the fundamental of $SL(5,\mathbb{R})$ and
suppressing the space-time form index on $A_{6,MN}$). Its orbit
corresponds to all components $A_{MM}$, i.e., those where the two
indices are equal. These are the five real roots contained in the
${\bf \overline{15}}$ representation and we recover the counting and
the same components coupling to supersymmetric tensor domain walls
as we did from the analysis of the WZ term. For the six-forms in the
${\bf \overline{40}}$, the highest weight is given by the component
$A^{1\,2,1}$. The highest-weight orbit consists then of all
components of $A^{MN,P}$ where $P=M$ or $P=N$. Therefore we arrive
again at the same criterion and counting as from the analysis of the
WZ term.

\section{Domain Walls and Gauged Supergravity}

In the previous section we reviewed the derivation of the elementary
half-supersymmetric domain walls in any maximal supergravity theory.
These are all the domain walls, i.e. $(D-2)$-branes, of a
$D$-dimensional  theory that admit a half-supersymmetric effective
action containing only one supersymmetric multiplet, that is either
a vector (for any $D$) or a tensor multiplet (for $D=7$). As we have
seen in the introduction for the case of Romans IIA in ten
dimensions, the presence of such a domain wall automatically induces
a deformation of the supergravity theory, that is the supergravity
theory is gauged. Maximal gauged supergravities in all dimensions
have been classified in \cite{deWit:2005hv,Samtleben:2005bp} in
terms of the so-called ``embedding tensor'', describing in a
U-duality covariant way how the gauge group embeds inside the
U-duality group. Denoting with $M_1$ the index of the U-duality
representation to which the 1-forms belong, and with $\alpha$ the
adjoint representation, the gauging leads to covariant derivatives
  \begin{equation}
  \partial_\mu  \mathbb{I}- g A_{\mu , M_1 } \Theta^{M_1}_\alpha t^\alpha \quad ,\label{covariantderivative}
\end{equation}
where $\Theta^{M_1}_\alpha$ is the embedding tensor and $t^\alpha$
are the generators of the U-duality group. Consistency with maximal
supersymmetry and gauge symmetry imposes constraints on the
U-duality representations the embedding tensor belongs to, and it
turns out that these representations are exactly conjugate to the
representations of the $(D-1)$-forms in the theory
\cite{Riccioni:2007au,Bergshoeff:2007qi,deWit:2008ta}. As we have
reviewed in the previous section, the elementary half-supersymmetric
$(D-2)$-branes are fewer than the number of U-duality components of
the corresponding $(D-1)$-form potentials, and in particular they
correspond to the highest-weight orbit of the highest-dimensional
irreducible representation (with the exception of the
seven-dimensional case, where there are vector domain walls in the
highest-weight orbit of the ${\bf \overline{40}}$ and tensor domain
walls in the highest-weight orbit of the ${\bf \overline{15}}$).
This means that these domain walls are associated to a particular
class of gauged supergravities, corresponding to an embedding tensor
having only non-zero components along these highest-weight orbits.
The aim of this section is to characterize these gauged theories.

As eq.~\eqref{covariantderivative} shows, the embedding tensor
groups together a subset of the abelian vectors of the ungauged
theory to form the adjoint of the gauge group, whose generators are
  \begin{equation}
  X^{M_1} =  \Theta^{M_1}_\alpha t^\alpha \quad , \label{Xgeneratorsgeneral}
  \end{equation}
  and whose commutation relations are given by
  \begin{equation}
[X^{M_1}, X^{N_1}] = f^{M_1 N_1}{}_{P_1} X^{P_1}\,,\ \ \ {\rm with}\
\ \  f^{M_1 N_1}{}_{P_1} = \big(X^{[M_1}\big)_{P_1}{}^{N_1]}
\end{equation}
where  $( X^{M_1} )_{P_1}{}^{N_1}$ is given as in
eq.~\eqref{Xgeneratorsgeneral} with the generators $t^\alpha$ acting
on the representation of the 1-forms. Consistency of the gauge
algebra not only imposes constraints on the representation of the
embedding tensor that we just mentioned (the so-called linear
constraints) but also the quadratic constraints
   \begin{equation}
  \Theta^{P_1}_\gamma ( X^{M_1} )_{P_1}{}^{N_1} = \Theta^{M_1}_\alpha \Theta^{N_1}_\beta f^{\alpha \beta}{}_\gamma \quad ,
  \end{equation}
where $f^{\alpha\beta}{}_{\gamma}$ are the structure constants of
the U-duality group, i.e. $[t^\alpha ,t^\beta ]=
f^{\alpha\beta}{}_{\gamma} t^\gamma$. Moreover, the 1-forms in
general also have to transform under the gauge parameter of the
2-forms $\Lambda_{1, M_2}$ as
\begin{equation}
\delta_{\Lambda_1} A_{1, M_1} = - g Z^{M_2}{}_{M_1} \Lambda_{1, M_2} \quad ,
\end{equation}
where the constants $Z^{M_2}{}_{M_1}$ satisfy the constraint
  \begin{equation}
Z^{M_2}{}_{M_1} \Theta^{M_1}_\alpha =0 \quad ,
\end{equation}
and we denote with $M_2$ the representation to which the 2-forms
belong. All these constraints guarantee that the gauging is
consistent. This means that for instance while a subset of the
abelian 1-forms of the ungauged theory  form the adjoint of the
gauge group, the remaining 1-forms  can either be uncharged with
respect to this gauge group or they are gauged away to give a mass
to some of the 2-forms in the theory. This gives rise to a hierarchy
of forms that continues all the way to the space-filling $D$-forms.

As a prototypical, although somewhat degenerate example, one can
consider the Romans mass deformation of the IIA theory. In this case
the embedding tensor vanishes, while $Z$ corresponds to the Romans
mass $m$. This means that the 1-form present in the massless theory
is gauged away to give mass to the 2-form, see Table
\ref{table:Romans}. In the rest of this section we will see how the
rearrangement of the degrees of freedom works explicitly in the
$D=9$, $D=8$ and $D=7$ gaugings. In each case we will select among
all the possible gaugings the ones that correspond to the
highest-weight orbit - that is the gaugings that admit elementary
domain-wall solutions. We will see how each orbit of gaugings
corresponds to a different rearrangement of the degrees of freedom,
and eventually we will point out what are the basic features of the
highest-weight orbit gaugings and how the degrees of freedom are
rearranged in these particular cases. Here we anticipate the result,
that is the highest-weight orbit gaugings are the deformations that
lead  to the minimal rearrangement of the degrees of freedom. Hence
we call these gaugings {\sl minimal}.

\subsection{The nine-dimensional gaugings}

In nine dimensions  the global symmetry is ${GL}(2,\mathbb{R})$, and
the 1-forms are $A_1$, $A_{1,a}$ in the ${\bf 1 \oplus 2}$. There is
also a doublet of 2-forms $A_{2,a}$ and a singlet 3-form $A_3$. The
gaugings of this theory have been classified in
\cite{Bergshoeff:2002nv} and then reconsidered using the embedding
tensor formalism in \cite{FernandezMelgarejo:2011wx}. The linear
constraints imply that the embedding tensor is $\Theta^a$,
$\Theta^{ab}$ belonging to the ${\bf 2 \oplus 3}$, while the
quadratic constraints are
  \begin{equation}
   \Theta^a \Theta^{bc} \epsilon_{ab} =0 \qquad \quad\Theta^{(a} \Theta^{bc)}= 0 \quad ,
\end{equation}
which imply that the two embedding tensors cannot be turned on together.

There is a single orbit of gauged theories associated to $\Theta^a$.
This corresponds to an $\mathbb{R}^+$ gauging. The 1-form $A_1$ is
gauged away by a shift $\Theta^a \Lambda_{1, a}$, where $\Lambda_{1,
a}$ is the parameter of the 2-form. If one takes $\Theta^1 =1$,
$\Theta^2 =0$, then $A_{2,1}$ is massive, while $A_{2,2}$ is gauged
away by the shift $\epsilon_{ab} \Theta^b \Lambda_2$, where
$\Lambda_2$ is the gauge parameter of the 3-form. Correspondingly,
the 3-form becomes massive.

Considering the $\Theta^{ab}$ gaugings, one has that the 1-forms
$A_{1,a}$ have the shift gauge symmetry $\epsilon_{ab} \Theta^{bc}
\Lambda_{1,c}$. There are three different orbits. Indeed, up to
${SL}(2,\mathbb{R})$ transformations, $\Theta^{ab}$ can be written
as ${\rm diag} (1,1)$, ${\rm diag}(1,-1)$ or ${\rm diag}(1,0)$. The
first two cases, corresponding to an ${SO}(2)$ and an ${SO}(1,1)$
gauging respectively, have the property that both 1-forms  $A_{1,a}$
are gauged away, leading to two massive 2-forms, while in the third
case only the 1-form $A_{1,2}$ is gauged away by the parameter
$\Lambda_{1,1}$. Correspondingly, only the 2-form $A_{2,1}$ is
massive and the other one remains massless. This last gauging, which
is  minimal because it gives the least amount of rearrangements of
the degrees of freedom, is exactly the highest-weight orbit gauging
corresponding to the two elementary half-supersymmetric domain walls
that we discussed in the previous section. The analysis of the
degrees of freedom for the $\Theta^{ab}$ gaugings  is summarized in
Table \ref{D=9gaugingstable}.

\begin{table}[h]
\hskip -1.5truecm
\begin{center}
{\small
\begin{tabular}{|c||c|c|c|c|c|c|}
\hline\rule[-1mm]{0mm}{6mm}
gauging &$A_1$ &$A_{1,1}$&$A_{1,2} $&$A_{2,1}$&$A_{2,2}$&$A_3$\\[.1truecm]
\hline \rule[-1mm]{0mm}{6mm}  $\Theta^{11} =1 \ \Theta^{22} = \pm 1$ &massive&gauged &gauged&massive&massive&massless\\[.05truecm]
\hline \rule[-1mm]{0mm}{6mm}$\Theta^{11} =1 \ \Theta^{22} =0$&massive&massless&gauged &massive&massless&massless\\[.05truecm]
\hline
\end{tabular}
}
\end{center}
 \caption{\sl The  $\Theta^{ab}$ gaugings  in nine dimensions. The last row corresponds to the minimal gauging leading to the elementary domain-wall solution.}\label{D=9gaugingstable}
\end{table}
\bigskip

\subsection{The eight-dimensional gaugings}
\label{sec:8dgauge}

In eight dimensions the symmetry is ${SL}(3,\mathbb{R}) \times
{SL}(2,\mathbb{R})$, and the propagating forms are the 1-forms
$A_{1, Ma}$ in the ${\bf ( \overline{3},2)}$, the 2-forms $A_{2}^M$
in the ${\bf (3,1)}$ and the 3-forms $A_{1,a}$ in the ${\bf (1,2)}$
which satisfy a self-duality condition. The most general gaugings of
this theory have been derived in
\cite{deRoo:2011fa,Dibitetto:2012rk}. The linear constraints select
the embedding tensors $\Theta_{MN}{}^a$ in the ${\bf
(\overline{6},2)}$ and $\Theta^{Ma}$ in the ${\bf (3,2)}$, with
quadratic constraints
  \begin{eqnarray}
& & \epsilon_{ab} \Theta^{Ma} \Theta^{Nb} =0 \nonumber \\
& & \Theta_{MN}{}^{(a} \Theta^{N b)} =0 \nonumber \\
& & \epsilon_{ab} \left(\epsilon^{MQR} \Theta_{QN}{}^a \Theta_{RP}{}^b + \Theta_{NP}{}^a \Theta^{Mb} \right)=0  \label{D=8quadraticconstraint}
\quad .
\end{eqnarray}

We know (see Table \ref{table1}) that the elementary domain walls
are associated to gaugings in the ${\bf (\overline{6},2)}$. More
precisely we know that there are 6 half-supersymmetric elementary
domain walls corresponding to the following gaugings
\begin{equation}
\begin{matrix}
\Theta_{11}{}^1 & & &\Theta_{11}{}^2\cr
&&&\cr
\Theta_{22}{}^1 & &  &\Theta_{22}{}^2\cr
&&&\cr
\Theta_{33}{}^1 & & &\Theta_{33}{}^2\cr
\end{matrix} \label{hwembeddingtensorD=8}
\end{equation}
It is easy to see that each embedding tensor in
eq.~\eqref{hwembeddingtensorD=8} satisfies the quadratic constraint
\eqref{D=8quadraticconstraint}.

\begin{table}[h]
\hskip -1.5truecm
\begin{center}
{\small
\begin{tabular}{|c||c|c|c|c|c|c|c|}
\hline\rule[-1mm]{0mm}{6mm}
gauging &$A_{1,11}$ &$A_{1,12}$&$A_{1,i1} $ & $A_{1,i2}$  &$A_{2}^1$&$A_{2}^i$&$A_{3,a}$\\[.1truecm]
\hline \rule[-1mm]{0mm}{6mm}$\Theta_{11}{}^1 =1$&massless&gauged&massive &massless&massive&massless& massless\\[.05truecm]
\hline
\end{tabular}
}
\end{center}
\caption{\sl The  minimal  gauging  in eight dimensions. The index
$i=2,3$ labels a global ${SL}(2,\mathbb{R})$ symmetry inside the
original ${SL}(3,\mathbb{R})$ which is preserved by the gauging.
}\label{D=8hwgaugingtable}
\end{table}

All the gaugings  in the  ${\bf (\overline{6},2)}$  can be obtained
reducing the 11-dimensional theory over group
manifolds~\cite{AlonsoAlberca:2003jq}. For such gaugings one can
show that imposing the quadratic constraints, and up to U-duality
transformations, one can always consider the index $a$ to be in the
1 direction, and $\Theta_{MN}^1$ in the diagonal form
  \begin{equation}
\Theta_{MN}{}^1 = {\rm diag}( \mathbb{I}_p , -\mathbb{I}_q , \mathbb{O}_r ) \qquad \quad p+q+r = 3
\end{equation}
leading to the gauge group \cite{AlonsoAlberca:2003jq,Bergshoeff:2003ri}
  \begin{equation}
  CSO (p,q,r) \quad .
\end{equation}
The structure constants of the gauge group are given by~\footnote{We
have dropped here the $SL(2,\mathbb{R})$ index $a=1$ on
$\Theta_{11}{}^1$ for ease of notation.}
\begin{equation}
f^{MN}{}_P = \epsilon^{MNQ}\Theta_{PQ} \quad ,
\end{equation}
leading to the following algebra
  \begin{equation}
  [ X^1 , X^2 ] = \Theta_{33} X^3 \ , \quad [ X^2 ,X^3 ] =  \Theta_{11} X^1 \ , \quad [ X^3 , X^1 ] = \Theta_{22} X^2 \quad .\label{ThetaalgebraD=8}
\end{equation}
For $p=3$, $q=r=0$ one gets the $SO(3)$ gauging of
\cite{Salam:1984ft}, and  one can obtain all the non-compact
gaugings by group contraction and/or analytic continuation from
$SO(3)$. The case $p=2$, $q=1$, $r=0$ is the $SO(2,1)$ gauging,
while for $r=1$ one can have $p=2$, $q=0$, which is an $ISO(2)$
gauging, or $p=1$, $q=1$, which is an $ISO(1,1)$ gauging. The
minimal  gaugings, associated to the elementary domain walls, have
$p=1$, $q=0$, $r=2$, corresponding to a gauge group $CSO(1,0,2)$.
Taking for instance $\Theta_{11}=1$ and $\Theta_{22}=\Theta_{33}=0$,
this is the three-dimensional Heisenberg algebra
  \begin{equation}
 [X^2 , X^3 ] = X^1 \quad , \qquad [X^1 , X^2 ] = [X^1 , X^3 ] =0 \quad .
\end{equation}
The gauge fields which acquire a mass by the Higgs mechanism are
$A_{1,i1}$, while the shift symmetry
$\Theta_{MN}{}^b\epsilon_{ab}\Lambda_1^N$ of $A_{1,Ma}$ gauges away
$A_{1,12}$ giving a mass to the 2-from $A_2^1$. All the other gauge
fields remain massless. In table \ref{D=8hwgaugingtable} we have
summarized this rearrangement. As in nine dimensions, this
rearrangement of the degrees of freedom with respect to the ungauged
theory is minimal and leads to the elementary domain walls.

\subsection{The seven-dimensional gaugings}
\label{sec:7dgauge}

The ungauged seven-dimensional maximal supergravity theory has
global U-duality symmetry ${SL}(5,\mathbb{R})$, and its
gaugings are determined by the embedding tensors  $\Theta_{MN,P}$ in
the ${\bf 40}$ and $\Theta^{MN}$  in the ${\bf 15}$. The propagating
forms are the 1-forms $A_{1,MN}$ in the ${\bf \overline{10}}$ and
the 2-forms $A_{2}^M$ in the ${\bf 5}$.

The elementary vector domain walls are associated to gaugings in the
highest-weight orbit of the ${\bf 40}$. These correspond to an
embedding tensor of the form $\Theta_{MN,M}$ for fixed $M$ and $N$,
giving in total 20 different gaugings. In general, for any embedding
tensor of the form $\Theta_{MN,P}= v_{[M} w_{N]P}$, with $v_M$ a
reference vector and with $w_{MN}$ symmetric, all the possible
gaugings have been classified in \cite{Samtleben:2005bp}, where it
was shown that  imposing the quadratic constraints, and up to
U-duality transformations, one can always consider $w_{MN}$ in the
diagonal form\,\footnote{Note that, given a fixed reference vector,
the indices of $w_{MN}$ effectively run from 1 to 4.}
  \begin{equation}
w_{MN} = {\rm diag}( \mathbb{I}_p , -\mathbb{I}_q , \mathbb{O}_r ) \qquad \quad p+q+r = 4
\end{equation}
leading to the gauge group $CSO (p,q,r)$. The particular case of the
highest-weight orbit corresponds to the case $p=1$, $q=0$, $r=3$
leading to the minimal gauging $CSO(1,0,3)$.
 In this case we can
consider as a representative of the highest-weight orbit the
component $\Theta_{12,1}$. This leads to the gauge algebra
\cite{deRoo:2006ms}
\begin{equation}
[X^i\,, X^j] = X^{ij}\,\hskip 1truecm i,j=3,4,5\,,
\end{equation}
where the indices  $i,j=3,4,5$ label the $SL(3, \mathbb{R})$ which
remains as a global symmetry of the deformed theory. The gauge
fields, acquiring a mass by eating three of the axions, are
$A_{1,ij}$. The shift symmetry of the 1-forms $\Theta_{MN,P}
\Lambda_1^P$ gauges away $A_{1,12}$. The vectors $A_{1,1i}$ and
$A_{1,2i}$ remain massless. The resulting rearrangement of the
degrees of freedom is summarized in Table \ref{D=7hwgauging40table}.
One can compare this table with Table 3 of  \cite{Cowdall:1996tw},
where a particular example of this orbit of gaugings was obtained as
a Scherk-Schwarz reduction from eight dimensions.  We stress that
among all possible gaugings in the ${\bf 40}$, the highest weight
one is the one that preserves the highest amount of global
symmetries.

\begin{table}[t]
\hskip -1.5truecm
\begin{center}
{\small
\begin{tabular}{|c||c|c|c|c|c|c|c|}
\hline\rule[-1mm]{0mm}{6mm}
gauging &$A_{1,ij}$ &$A_{1,12}$&$A_{1,1i} $ & $A_{1,2i}$  &$A_{2}^1$&$A_{2}^2$&$A_{2}^i$\\[.1truecm]
\hline \rule[-1mm]{0mm}{6mm}$\Theta_{12,1} =1$&massive&gauged&massless &massless&massive&massless& massless\\[.05truecm]
\hline
\end{tabular}
}
\end{center}
\caption{\sl The  minimal gauging in the ${\bf 40}$  in seven
dimensions. The index $i=3,4,5$ labels a global ${SL}(3,\mathbb{R})$
symmetry inside the original ${SL}(5,\mathbb{R})$ which is preserved
by the gauging. }\label{D=7hwgauging40table}
\end{table}

We now consider the gaugings associated to the elementary tensor
domain walls. These gaugings belong to the highest-weight orbit in
the ${\bf 15}$, corresponding to an embedding tensor $\Theta^{MN}$
of the form $\Theta^{MM}$ for fixed $M$. In general, all possible
gaugings in the ${\bf 15}$ have been classified in
 \cite{Samtleben:2005bp}, where it was shown that  imposing the quadratic
constraints, and up to U-duality transformations, one can always
consider $\Theta^{MN}$
in the diagonal form
  \begin{equation}
\Theta^{MN} = {\rm diag}( \mathbb{I}_p , -\mathbb{I}_q , \mathbb{O}_r ) \qquad \quad p+q+r = 5 \quad ,
\end{equation}
which again results in  the gauge group $CSO (p,q,r)$. The minimal
gaugings correspond to $p=1$, $q=0$, $r=4$, leading to the gauge
group $CSO(1,0,4)$. Considering for instance $\Theta^{11}$ as the
only non-vanishing component, one can see that the gauge vectors,
acquiring a mass by the Higgs mechanism gauging away four axions,
are $A_{1, 1i}$, with the index $i$ now labeling the directions
$2,3,4,5$ in the fundamental of ${SL}(5,\mathbb{R})$. These indices
label the global $SL(4,\mathbb{R})$ symmetry which is preserved by
this gauging. As in the previous case, this is the highest amount of
global symmetries that is preserved by any possible gauging in the
${\bf 15}$. The other six vectors $A_{1, ij}$ remain massless, while
one of the 2-forms is gauged away by the shift symmetry $\Theta^{MN}
\Lambda_{2,N}$, where $\Lambda_{2,M}$ are the gauge parameters of
the 3-forms $A_{3,M}$ that in the ungauged theory are dual to the
2-forms. Correspondingly, one of these 3-forms acquires a mass. This
is consistent with the counting of the degrees of freedom because in
the gauged theory this 3-form satisfies a massive self-duality
condition. The list of all the degrees of freedom for this gauging
is summarized in Table \ref{D=7hwgauging15table}. Again, one can
compare this table with Table 4 of \cite{Cowdall:1996tw}.

\begin{table}[h]
\hskip -1.5truecm
\begin{center}
{\small
\begin{tabular}{|c||c|c|c|c|c|}
\hline\rule[-1mm]{0mm}{6mm}
gauging &$A_{1,1i}$ &$A_{1,ij} $  &$A_{2}^1$&$A_{2}^i$&$A_{3,1}$\\[.1truecm]
\hline \rule[-1mm]{0mm}{6mm}$\Theta^{11} =1$&massive&massless&gauged &massless&massive\\[.05truecm]
\hline
\end{tabular}
}
\end{center}
\caption{\sl The minimal  gauging in the ${\bf 15}$  in seven
dimensions. The index $i=2,3,4,5$ labels a global
${SL}(4,\mathbb{R})$ symmetry inside the original
${SL}(5,\mathbb{R})$ which is preserved by the gauging.
}\label{D=7hwgauging15table}
\end{table}
\bigskip

In lower dimensions, all these results continue to hold, namely:
\begin{itemize}
\item the elementary domain walls are solutions of the gauged supergravity
theories obtained by taking the embedding tensor to take values in
the highest-weight orbit of the relevant representation;

\item an embedding tensor taking values in  the highest-weight orbit  satisfies the quadratic constraints and leads to a minimal gauging;

\item this gauging preserves the highest amount of global symmetries among all possible gaugings in the same representation.

\end{itemize}

In the next section we will discuss domain-wall solutions, and show
that one can obtain (non-elementary) half-supersymmetric domain-wall
solutions also for gaugings that are not in the highest-weight
orbit.

\section{Domain-wall Solutions}

In the previous two sections we have classified the elementary
supersymmetric domain walls of toroidally compactified IIA/IIB
string theory and specified the minimal gauged supergravity theories
they correspond to. In this section we wish to investigate the
domain-wall solutions of generic maximally supersymmetric gauged
supergravities, including the non-minimal ones. We already discussed
the 10D case in the introduction, in which case there is only one
elementary D8-brane which is a solution of massive IIA supergravity.
In the next two subsections  we will discuss the situation both from
a supergravity as well as from an $E_{11}$ point of view.

\subsection{Domain-wall solutions and supergravity}

Below we will discuss domain-wall solutions of gauged supergravity theories  in 9D, 8D and 7D, respectively.
\bigskip

\noindent {\sl D=9:}\ \  In 9D there are two distinct embedding
tensors. One is  a doublet $\Theta^a\ (a=1,2)$ and the other is a
triplet $\Theta^i\ (i=1,2,3)$\,\footnote{In the previous section we
have denoted this embedding tensor as $\Theta^{ab}$, symmetric in
$ab$. The relation between this object and $\Theta^i$ is
$\Theta^{ab} = \Theta^i t_i^{ab}$, where $t_i^{ab}$ are the
$SL(2,\mathbb{R})$ generators. This leads to the identifications
$\Theta^{11} = \Theta^+$, $\Theta^{22} = \Theta^-$ and $\Theta^{12}
= \Theta^3$ (see eq. \eqref{theta+theta-theta3}).} of the U-duality
group $GL(2,\mathbb{R})$. The projection operator for a domain wall
is given by
\begin{equation}\label{BPS9D}
\tfrac{1}{2}\big(1 \pm \gamma_y\big)\epsilon_0=0\,,
\end{equation}
where $\epsilon_0$ is a constant spinor and $y$ denotes the
transverse direction. From an investigation of the Killing spinor
equations it follows that the gauged supergravities corresponding to
the $\Theta^a$ gaugings do not have any half-super\-sym\-me\-tric
domain-wall solutions \cite{Bergshoeff:2002nv}. This in accordance
with the fact that only the highest-dimensional $(D-1)$-form
potentials couple to supersymmetric domain walls.

The $\Theta^i$ gaugings do allow half-supersymmetric domain-wall
solutions which have been extensively investigated in
\cite{Bergshoeff:2002mb}. Choosing a lightcone direction
\begin{equation}
\Theta^i = (\Theta^+,\Theta^-, \Theta^3) = (1,0,0) \label{theta+theta-theta3}
\end{equation}
one obtains a minimal $\mathbb{R}^+$-gauging (see the previous
section). This minimal gauging allows an elementary domain-wall
solution which can be oxidized to the IIB D7-brane solution.
Choosing the other lightcone direction, i.e.~$\Theta^i= (0,1,0)$, a
similar domain-wall solution and potential is obtained. This one
oxidizes to the S-dual of the IIB D7-brane solution.

Choosing a non-light-cone direction, i.e.
\begin{equation}
\Theta^i  = (0,0,1)
\end{equation}
one obtains a non-minimal $SO(1,1)$-gauging. The corresponding
gauged theory admits a supersymmetric domain wall that is not an
elementary brane. Instead, this solution describes the threshold
bound state of two elementary domain walls.

\bigskip

\noindent {\sl D=8:}\ \ As we have reviewed in the previous section,
in 8D we have two embedding tensors. One embedding tensor
$\Theta_{MN}{}^a$\ $(M=1,2,3; a=1,2)$ transforms in the ${\bf
(\overline{6},2)}$ of the U-duality group $SL(3,\mathbb{R})\times
SL(2,\mathbb{R})$. The other embedding tensor $\Theta^{Ma}$
transforms in the ${\bf (3,2)}$ representation. The supersymmetric
domain walls of 8D maximal gauged supergravity have been extensively
discussed in \cite{AlonsoAlberca:2003jq,Bergshoeff:2003ri}. As
expected, one finds that the lowest dimensional ${\bf ( 3,2)}$
representation does not lead to any supersymmetric domain-wall
solutions. The highest-dimensional ${\bf (\overline{6},2)}$
representation leads to 6 half-supersymmetric domain-wall solutions.
The minimal gaugings corresponding to these elementary domain walls
have been discussed in subsection~\ref{sec:8dgauge}.

The elementary domain walls can be obtained as truncations of the
general domain-wall solutions given in
\cite{AlonsoAlberca:2003jq,Bergshoeff:2003ri} which contain three
independent harmonic functions. The results of
\cite{AlonsoAlberca:2003jq,Bergshoeff:2003ri} show that there are
many more supersymmetric domain-wall solutions corresponding to more
general gaugings. For instance, the $ISO(2)$-gaugings or
$SO(3)$-gaugings
\begin{equation}
ISO(2):\  \Theta_{11}{}^1 = \Theta_{22}{}^1=1\,;\hskip 1.5truecm
SO(3):\  \Theta_{11}{}^1 = \Theta_{22}{}^1=\Theta_{33}{}^1=1
\end{equation}
allow supersymmetric domain walls with two and three independent
harmonic functions, respectively. These  supersymmetric domain-wall
solutions do not correspond to elementary domain walls. Instead,
they should be viewed as threshold bound states of the elementary
domain walls. The 11D origin of these threshold bound states has
been discussed in \cite{Bergshoeff:2003ri}.

\bigskip

\noindent {\sl D=7:}\ \
In 7D there are two embedding tensors. One embedding tensor $\Theta_{MN,P}$  transforms in the ${\bf 40}$ of the ${SL}(5,\mathbb{R})$
U-duality symmetry.
The other embedding tensor $\Theta^{MN}$ transforms in the ${\bf 15}$ representation.
What is special in 7D is that domain walls are 5-branes and there are two types of them: domain walls with
worldvolume vector multiplets and domain walls with worldvolume tensor multiplets. The $\Theta_{MN,P}$ lead to
gaugings that allow vector domain walls, while  the lower-dimensional embedding tensor  $\Theta^{MN}$ also
allows supersymmetric domain-wall solutions that have tensor instead of vector multiplets on the worldvolume.

A systematic investigation of the supersymmetric domain-wall solutions of 7D maximal gauged supergravity has not been performed so far.
Following our general analysis, we expect that the minimal gaugings discussed in subsection~\ref{sec:7dgauge} lead to 20 elementary vector
domain-wall solutions and 5 tensor domain-wall solutions.
It is interesting to consider  the $\Theta^{MN}$ and $\Theta_{MN,P}$ gaugings together. Defining
\begin{equation}
s = {\rm rank}\, \Theta_{MN,P}\,\hskip 2truecm t = {\rm rank}\, \Theta^{MN}
\end{equation}
where $\Theta_{MN,P}$ is understood as a rectangular $10 \times 5$
matrix, one finds the following re-arrangement of the 100 degrees of
freedom carried by the different 1-forms, 2-forms and 3-forms
\cite{Samtleben:2005bp}, see Table \ref{combined}.\,\footnote{ We do
not indicate how many of the 1-forms are massless or massive. This
would require a more detailed analysis of the gauge transformations
including the axions.}
 For
$(s,t)=(1,0)$ and $(s,t)=(0,1)$ this table reproduces the results of
Tables \ref{D=7hwgauging40table} and \ref{D=7hwgauging15table},
respectively. Note that the quadratic constraint ensures that
$s+t\leq 5$.

\begin{table}[h]
\hskip -1.5truecm
\begin{center}
{\small
\begin{tabular}{|c||c|c|c|c|}
\hline\rule[-1mm]{0mm}{6mm} form & 1-forms &2-forms& 2-forms & s.d.
3-forms \\ [.1truecm] \hline \rule[-1mm]{0mm}{6mm}
mass&massless/massive&massless &massive&massive \\ [.1truecm] \hline
\rule[-1mm]{0mm}{6mm}
\# &$10 - s$&$5 - s - t$&$s$&$t$\\
\hline
\end{tabular}
}
\end{center}
 \caption{\sl The  minimal gauging in the ${\bf 40}\oplus {\bf 15}$  in seven dimensions.
}\label{combined}
\end{table}

Unlike the previous cases, since we are now dealing with two {\sl
different} BPS conditions (see also section~\ref{sec:CC}), we expect
this to lead to domain walls with {\sl less} supersymmetry, i.e.~1/4
supersymmetric domain walls. Indeed such 1/4-supersymmetric domain
wall solutions are discussed  in \cite{Cowdall:1996tw}, where it is
also shown that a particular representative of these gauged
supergravity theories is obtained as a generalized Scherk-Schwarz
reduction where two different axions acquire a linear dependence on
the compactified coordinate. By reduction we expect similar
1/4-supersymmetric domain-wall solutions to occur in $D<7$
dimensions.

\subsection{Domain-wall solutions and $E_{11}$}

We now explain some of the algebraic mechanics of domain-wall
solutions from a one-dimensional $E_{11}$ coset model point of view.
A one-dimensional coset model based on $E_{11}$ with the so-called
temporal involution was presented in~\cite{Englert:2003py} based
on~\cite{Damour:2002cu}. The coset model is given by a map
\begin{align}
\mathcal{V} \,:\, \mathbb{R} \to E_{11} / K^*(E_{11}),\quad\quad \xi\mapsto \mathcal{V}(\xi),
\end{align}
where $K^*(E_{11})$ is the subgroup of $E_{11}$ fixed by the
temporal involution~\cite{Englert:2003py}. This involution is such
that $K^*(E_{11})$ contains the Lorentz group $SO(10,1)$ rather than
the compact rotation group $SO(11)$ that one would obtain with the
standard Chevalley involution on $E_{11}$.\footnote{Using the
temporal involution also groups together possible different choices
of space-time signature~\cite{Keurentjes:2004bv,Keurentjes:2004xx}.}
The action for the coset model is the standard one for null geodesic
motion~\cite{Damour:2002cu,Englert:2003py,Damour:2004zy}
\begin{align}
\label{sigmaact}
S = \int d\xi  \frac{1}{2n} \langle \mathcal{P}|\mathcal{P}\rangle ,
\end{align}
with the Killing bilinear form $\langle\cdot|\cdot\rangle$ and where
the lapse $n(\xi)$ is needed for reparametrization invariance of the
geodesic and
\begin{align}
\mathcal{P} = \frac12\left( \partial_\xi \mathcal{V} \mathcal{V}^{-1} + \left( \partial_\xi \mathcal{V} \mathcal{V}^{-1} \right)^\# \right),
\end{align}
where $(\cdot)^\#$ denotes the application of minus the temporal
involution. Thinking of  $\mathcal{V}$ as a matrix, the operation
$\mathcal{V}^\#$ can be though of as transposing $\mathcal{V}$ and
multiplying by the (generalization) of the Lorentz metric from left
and right. $\mathcal{P}$ is then (roughly) the symmetric part of the
Maurer-Cartan form.

We note that the domain walls are the perfect playground for the
one-dimensional sigma model since domain walls depend only on one
transverse direction which can be identified with the geodesic
parameter $\xi$. This can be used to explicitly determine the
space-time solution from a solution of the one-dimensional sigma
model once a map between the coset variables and space-time
variables is known~\cite{Englert:2003py,Kleinschmidt:2005gz}.

For our elementary branes we are interested in configurations that
use only a single $E_{11}$ step operator $E_\alpha$ (for a given
root $\alpha$) out of the infinitely
many~\cite{Englert:2003py,West:2004st,Cook:2004er,Kleinschmidt:2005gz},
see also~\cite{Cook:2009ri,Houart:2009ya}. We parametrize the coset
element then as
\begin{align}
\mathcal{V} = e^{\phi h_{\alpha}}e^{ A E_{\alpha}},
\end{align}
where $E_{\alpha}$ is the single step operator and $h_\alpha$ is its
associated element in the Cartan subalgebra. We assume
$E_{\alpha}+(E_{\alpha})^\#$ to be associated with a compact
direction. The reason for this that we are interested in brane
solutions rather than cosmological
solutions~\cite{Englert:2004ph,Kleinschmidt:2005gz}. In this
$SL(2,\mathbb{R})$ subsector (for real $\alpha$) all coset equations
of motion are solved by~\cite{Englert:2004ph,Kleinschmidt:2005gz}
\begin{align}
\label{sl2sol}
e^{2\phi(\xi)} = H(\xi) \quad\text{and}\quad A = \pm H^{-1}
\end{align}
for any harmonic function $H(\xi)$. The variable $A$ is cyclic in the coset dynamics and generates an effective potential for the scalar field $\phi$ similar to the scalar potential of the associated gauged supergravity.
For imaginary roots $\alpha$ the
solution looks very different~\cite{Kleinschmidt:2005gz}.

Using the known standard dictionary for supergravity solutions~\cite{Damour:2002cu},
one expects the scalar field to be related to a combination of the dilaton fields in
maximal supergravity in $D$ dimensions. Since the metric is also related to exponentials
of $\phi$, a constant $\phi$ means in fact a flat space solution which is clearly not a
domain wall. Since the dictionary is not established for imaginary roots the exact general
space-time interpretation of the solution to the geodesic model is not known in that case.
In the null case a thorough investigation was carried out in~\cite{Houart:2011sk}, showing
that the corresponding space-time solution is not supersymmetric.\\

{\noindent {\sl Example 1:}} The D$8$-brane of massive type IIA can
be described in this language~\cite{West:2004st,Englert:2007qb}. For
this one requires the root vector $\alpha$ that is related to the
$9$-form and the relation of $\alpha$ to a basis of metric
components and dilaton. The detailed change of basis can be found
for example
in~\cite{West:2004st,Damour:2002fz,Englert:2007qb,Englert:2003zs}.
Here, we only give the result. The root vector $\alpha$ for a
D$8$-brane (in directions $1,\ldots,9$) is
\begin{align}
\label{d8rt}
\alpha  = \left( \frac9{16},\frac1{16},\frac1{16},\frac1{16},\frac1{16},\frac1{16},\frac1{16},\frac1{16},\frac1{16},\frac1{16};\frac54 \right) = \left(p_i;p_\phi\right).
\end{align}
From this root vector one obtains the diagonal metric and
dilaton by taking $g_{ii}=\epsilon_i H^{2p_i}$ and $e^\phi = H^{p_\phi}$,
where $\epsilon_i=\pm 1$ is the signature of the $i$th direction.
This reproduces the Einstein frame metric (\ref{D8met}). The form field from (\ref{sl2sol}) also is correct compared to (\ref{D8met}).\\

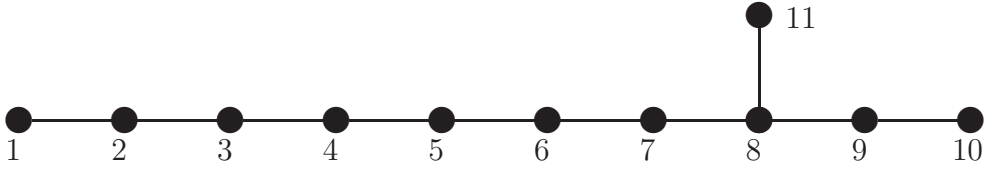
\begin{figure}[t!]
\centering
\begin{picture}(380,60)
\put(5,-5){$1$}
\put(45,-5){$2$}
\put(85,-5){$3$}
\put(125,-5){$4$}
\put(165,-5){$5$}
\put(205,-5){$6$}
\put(245,-5){$7$}
\put(285,-5){$8$}
\put(325,-5){$9$}
\put(363,-5){$10$}
\put(300,45){$11$}
\thicklines
\multiput(10,10)(40,0){10}{\circle*{10}}
\multiput(15,10)(40,0){9}{\line(1,0){30}}
\put(290,50){\circle*{10}} \put(290,15){\line(0,1){30}}
\put(290,10){\circle*{10}}
\end{picture}
{\caption{\label{fig:e11dynk}\sl Dynkin diagram of $E_{11}$ with labelling of nodes.}}
\end{figure}

{\noindent {\sl Example 2:}} As another example, we consider the
intersection of two $1/2$-BPS domain walls in $D=7$. As there are
now several dilatonic scalars involved, we refrain from giving the
full metric but only indicate the $E_{11}$ roots. Expanded on a
basis of simple roots labeled according to the $E_{11}$ Dynkin
diagram of figure~\ref{fig:e11dynk} one can choose a $1/2$-BPS
vector domain wall with corresponding real root $\alpha_{\text v}$
and a $1/2$-BPS tensor domain wall with corresponding root
$\alpha_{\text t}$ where
\begin{align}
\label{14bps}
\alpha_{\text v} &= \alpha_2+ 2\alpha_3+3\alpha_4+4\alpha_5+5\alpha_6+6\alpha_7+6\alpha_8+3\alpha_9+\alpha_{10}+3\alpha_{11},&\nonumber\\
\alpha_{\text t} &= \alpha_2+ 2\alpha_3+3\alpha_4+4\alpha_5+5\alpha_6+6\alpha_7+6\alpha_8+4\alpha_9+2\alpha_{10}+2\alpha_{11}.&
\end{align}
One can check that these roots are Cartan orthogonal, i.e., they
satisfy $\alpha_{\text v}\cdot\alpha_{\text t}=0$ with respect to
the Cartan inner product. They therefore correspond to an orthogonal
intersection of $1/2$-BPS branes~\cite{Englert:2004it}. That the
intersecting solution is $1/4$-BPS can be inferred from the analysis
of the BPS conditions of the following section. We  remark that,
from an M-theory perspective, the configuration above corresponds to
the intersection of a KK$7$-monopole with an M$5$-brane. Bound
states viewed from an algebraic perspective were also discussed
in~\cite{Cook:2009ri,Houart:2009ya}.

\section{Domain Walls and Central Charges}
\label{sec:CC}

In this section, we study the BPS conditions satisfied by elementary
supersymmetric domain walls. We will do this  first from a field
theory perspective and, next,  from an $E_{11}$ point of view.

\subsection{One central charge, many domain walls}

It is well-known that there is a 1-1 relation between the
half-super\-sym\-me\-tric branes of maximal supergravity with more
than or equal to three transverse directions and the central charges
in the supersymmetry algebras with 32 supercharges \cite{de
Azcarraga:1989gm,Townsend:1997wg}. It is less obvious what the
precise relation is for the branes with less than three transverse
directions. This is due to the fact that these branes are not
asymptotically flat and hence one cannot define charges for these
objects. Nevertheless, one expects a relation, be it not 1-1,
between  the BPS conditions of the different non-standard branes and
the central charges. Indeed, for branes with two transverse
directions, i.e.~``defect branes'', it has been found that this
relation is always 1-2, i.e., each BPS condition corresponds to {\sl
two} defect branes \cite{Bergshoeff:2011se}. These two defect branes
are related by an S-duality to each other.

It is instructive to consider, as an example, the
half-supersymmetric 7-branes of IIB string theory and  the central
charges of the 10D IIB superalgebra. This algebra has a single
3-form central charge $Z_{abc}$ which is a singlet under the
$R$-symmetry group $SO(2)$. On the one hand there is a 1-1 relation
between this central charge and the D3-brane which is a singlet
under the U-duality group $SL(2,\mathbb{R})$. Alternatively, one may
consider the dual central charge ${\tilde Z}_{a_1\cdots a_7}$ and
its relation to the 7-branes of IIB string theory.

To be more specific, consider a 7-brane extended in the directions
$x^1\dots x^7$. We define the complex transverse coordinate
$z=x^8+ix^9$. Since we consider supersymmetric solutions the Killing
spinor equations must be satisfied. The most general  solution to
these equations is given by \cite{Greene:1989ya,Bergshoeff:2006jj}
\begin{eqnarray}
ds^2 & = & -dt^2+d{{\vec x}_{7}}^{\;2}+({\rm Im}\tau)\vert
f\vert^2 dz d\bar
z\,,\label{metric}\nonumber\\[.2truecm]
\tau & = & \tau(z)\,,\quad f=f(z)\,,\label{tau} \\[.2truecm]
\epsilon & = & \left(f/\bar f\right)^{\!\!1/4}\!\!\epsilon_0\,,\nonumber
\end{eqnarray}
\noindent where  $\tau=\chi + ie^{-\phi}$  is the axion-dilaton,
$f(z)$ is a holomorphic function and  $\epsilon_{0}$ is a constant
spinor which satisfies
$\gamma_{{\underline{z}}^{*}}\epsilon_{0}=0$.\,\footnote{We use
complex notation in which  $\epsilon$ can be written as
$\epsilon=\epsilon_1+i\epsilon_2$ where $\epsilon_1$ and
$\epsilon_2$ are two Majorana-Weyl spinors. We take the chirality of
$\epsilon$ to be  negative, i.e.~$\gamma_{11}\epsilon=-\epsilon$.
The underbar in $\underline{z}$ indicates that this is a flat
index.} Under $SL(2,\mathbb{R})$ the holomorphic functions
$\tau(z)\,,f(z)$ and $\epsilon(z)$  transform as
\begin{equation}
\tau\rightarrow\Lambda\tau \equiv \frac{a\tau+b}{c\tau+d}\,,\hskip .8truecm  f(z) \rightarrow (c\tau(z)+d)f(z)\,, \hskip .8truecm \epsilon \rightarrow e^{i\varphi}\epsilon\,,
\end{equation}
where
\begin{equation}
\Lambda= \left(
\begin{array}{cc}
a & b \\
c & d \\
\end{array}
\right)\in SL(2,\mathbb{R})\, \hskip .7truecm  {\rm and}\hskip .7truecm  \varphi = \tfrac{1}{2}{\rm arg} (c\tau +d)\,.
\end{equation}
More precisely, this means that $\tau(z)$ transforms under $PSL(2,\mathbb{R})$ and that $\epsilon(z)$ transforms under the
double cover of $SL(2,\mathbb{R})$.

From the general supersymmetric configuration \eqref{metric} we deduce that {\sl all} solutions satisfy the following
$SL(2,\mathbb{R})$-invariant BPS condition:
\begin{equation}\label{BPSc}
\gamma_{{\underline{z}}^{*}}\epsilon=0\,.
\end{equation}
From \cite{Bergshoeff:2011se} we know that there are two elementary
supersymmetric 7-branes. Using real notation $SO(2,1) \simeq SL(2,
\mathbb{R})$  they correspond to the two lightcone directions of
$SO(2,1)$. These branes are the D7-brane and its S-dual. This
confirms the two-fold degeneracy of the BPS condition \eqref{BPSc}
mentioned in \cite{Bergshoeff:2011se}. From the general analysis
above we deduce that there is a third half-supersymmetric 7-brane
solution, corresponding to the third non-lightcone direction,  that
is not elementary, i.e.~it has no supersymmetric WZ term. This
7-brane  solution describes  a threshold bound state of a D7-brane
and an S-dual $ {\rm D7}$-brane. The 10D 7-brane situation is
generic for all defect branes in $D\le 10$ dimensions
\cite{Bergshoeff:2011se}. There are always twice as many defect
branes as central charges since each defect brane and its S-dual
have the same BPS projection operator.

We now wish to investigate whether a similar, not necessarily 1-1,
relation like we just found for the defect branes also  holds
between the central charges of the algebras with 32 supercharges and
the elementary domain walls studied in this work. The central
charges corresponding to supersymmetric domain walls are the duals
$Z_{a_1\cdots a_{D-2}}$ of the 2-form central charges $Z_{ab}$ for
$4 \le D \le 10$ dimensions while they are the 1-form central
charges $Z_a$ for $D=3$ dimensions. These central charges transform
as representations of the $R$-symmetry group, see Table
\ref{table:centralcharges}, and we study them first for the standard
$2$- and $1$-branes that couple to them.

We expect a 1-1 relation between these central charges and branes
with three or more transverse directions. We wish to verify this 1-1
relation by comparing, for each dimension, the central charges with
the corresponding 2-branes in $D\geq 6$. Starting with 10D we note
that  in the 10D IIA theory there is a single 2-form central charge
and, correspondingly, a single D2-brane. In 9D the relevant brane is
the D2-brane compactified over one of its transverse directions. In
8D there is a doublet of Dirichlet 2-branes which transform as a
chiral spinor representation of the T-duality group
$SL(2,\mathbb{R})\times SL(2,\mathbb{R})$. In 7D there is a
4-component T-duality  spinor of Dirichlet 2-branes and a singlet
solitonic 2-brane. The second singlet given in Table
\ref{table:centralcharges} is a Kaluza-Klein monopole.\,\footnote{As
far as the central charge and BPS condition is concerned a KK
monopole in $D$ dimensions behaves as a $(D-5)$-brane, i.e.~in $D=7$
dimensions it behaves as a 2-brane.} This adds up to a total of 6
branes corresponding to the 6 2-form central charges indicated in
the table. In 6D there is a 8-component T-duality spinor of
Dirichlet 2-branes and a 8-component T-duality vector of solitonic
2-branes adding up to a total of  16 branes corresponding to the 16
2-form central charges in 6D.

We now consider the relation between the 2-form charges and the
2-branes for $ D=5$ and $D=4$, as well as the relation between
1-form central charges and 1-branes in 3D. In 5D 2-branes have {\sl
two} transverse directions, i.e.~they are defect branes. We
therefore expect a 1-2 relation in this case. Indeed, there is a
16-component chiral T-duality spinor of Dirichlet 2-branes, a
40-dimensional orbit of solitonic 2-branes within the 45-dimensional
adjoint representation of the T-duality group and a 16-component
anti-chiral T-duality spinor of 2-branes with $\alpha=-3$. This adds
up to a total number of 72 defect branes which, as expected, is
twice the number of 5D 2-form central charges. In 4D 2-branes {\sl
are} domain walls and their total number is indicated in the last
column of Table \ref{table:centralcharges}. Finally, in 3D domain
walls are 1-branes and their total number is also given in the last
column.

Returning to domain walls we can read off from Table
\ref{table:centralcharges} some general patterns although they are
not as clean as in the case of the defect branes. We observe that
for $5\le D \le 10$ dimensions the number $n_{{\rm DW}}$ of
supersymmetric domain walls
 with a worldvolume vector multiplet
is  $d+1$ times the number $n_{{\rm Z}}$ of corresponding 2-form
central charges, i.e.\footnote{Note that in 7D the central charges
occur in two different representations. This is related to the fact
that in 7D there are two types of domain walls, with a different
worldvolume content.  There are 20 domain walls with a worldvolume
vector multiplet for which eq.~\eqref{rule} applies, i.e.~$n_{\rm
DW} = 4 n_{\rm Z}$ and there are five domain walls with a
worldvolume tensor multiplet for which we have $n_{\rm DW} = 5
n_{\rm Z}$.\label{footnote:domainwalls}}
\begin{equation}\label{rule}
n_{\rm DW} = (d+1)\,n_{\rm Z}\,,\hskip 1truecm 5 \le D \le 10\,.
\end{equation}
Here, $d=10-D$. For $D=4$ and $D=3$ the relations are
$n_{\rm DW} = 8\,n_{\rm Z}$ and
$n_{\rm DW} = 16\,n_{\rm Z}$, respectively.
 Below we will discuss, for each dimension  $6\le D\le 10$ separately, starting with ten dimensions,
 how this degeneracy  fits with our results on the embedding tensor obtained in section 3.

\bigskip

\begin{table}[h]
\hskip -.4truecm
\begin{tabular}{|c|c|c|c|c|c|}
\hline \rule[-1mm]{0mm}{6mm}
$D$&$H$&$n=1$&$n=2$&2-branes&$n_{\rm DW}$\\[.1truecm]
\hline \rule[-1mm]{0mm}{6mm} IIA &1&&${\bf 1}$&${\bf 1}_{-1}$&1 \\[.05truecm]
\hline \rule[-1mm]{0mm}{6mm} 9&$SO(2)$&&{\bf 1}&${\bf 1}_{-1}$&2 \\[.05truecm]
\hline \rule[-1mm]{0mm}{6mm} 8&$SO(3)\times SO(2)$&&${\bf (1,2)}$&${\bf (1,2)}_{-1}$&6\\[.05truecm]
\hline \rule[-1mm]{0mm}{6mm} 7&$Sp(4)$&&${\bf 5}+{\bf 1}$&(${\bf 4}_{-1}$ + ${\bf 1}_{-2}$) + 1&20+5 \\[.05truecm]
\hline \rule[-1mm]{0mm}{6mm} 6&$Sp(4) \times  Sp(4)$&&${\bf (4,4)}$&$({\bf 8_{\rm S}})_{-1} + ({\bf 8_{\rm C}})_{-2}$& 80\\[.05truecm]
\hline \rule[-1mm]{0mm}{6mm} 5&$Sp(8)$&&{\bf 36}&${\bf 16}_{-1} + 40\subset {\bf 45}_{-2} + {\bf {\overline {16}}}_{-3}$&216 \\[.05truecm]
\hline \rule[-1mm]{0mm}{6mm} 4&$SU(8)$&&${\bf 36}^+ + {\bf \overline {36}}^-$&&576 \\[.05truecm]
\hline \rule[-1mm]{0mm}{6mm} 3&$SO(16)$&${\bf 135}$&&&2160\\[.05truecm]
\hline
\end{tabular}
 \caption{\sl This table indicates the 1-form and 2-form central charges that are related to supersymmetric domain walls. The number $n$
 in the top row indicates the rank of the central charge.
The $R$-symmetry group $H$ is indicated in the second column. The
next two columns indicate the representations of $H$ according to
which  the 1-form and 2-form central charges transform. The fifth
column gives the T-duality representations of the 2-branes (not the
domain walls) associated to the 2-form central charges. The
sub-index indicates the $\alpha$-value of these branes. The singlet
in 7D, without a sub-index, corresponds to a Kaluza-Klein monopole.
We have left the entries for the 4D and 3D cases empty since 4D
2-branes and 3D 1-branes are domain walls and as such are
represented in the last column.  This last column gives the total
number of half-supersymmetric domain walls corresponding to the
duals of the 1-form and 2-form central charges. Their
$\alpha$-values are given in Table \ref{table1}.
  \label{table:centralcharges}}
\end{table}

\bigskip

\noindent {\sl D=10}\,:\ \ This case has been discussed extensively
in the introduction. There is a single mass parameter $m$ which is a
massive deformation (without any gauging)  of IIA supergravity. The
D8-brane is the half-supersymmetric domain-wall solution of this
massive IIA supergravity theory. Its BPS condition corresponds to
the  dual of the  2-form central charge in the IIA supersymmetry
algebra. In this case we find a 1-1 relation between the dual of the
2-form central charge and the half-supersymmetric domain wall of the
theory.
\bigskip

\noindent {\sl D=9}\,:\ \ In the previous section we have seen that
the $\Theta^i\ (i=1,2,3)$ gau\-gings allow two elementary
half-super\-sym\-me\-tric domain-wall solutions with the same BPS
condition \eqref{BPS9D}. These solutions correspond to the two
lightcone directions of $SL(2,\mathbb{R}) \simeq SO(2,1)$. Comparing
with Table \ref{table:centralcharges} we conclude that there is a
two-fold degeneracy: there are 2 half-supersymmetric elementary
domain walls corresponding to the single 7-form dual of the 2-form
central charge.

\bigskip

\noindent {\sl D=8}\,:\ \ In subsection 3.2 we found that there are
three different minimal gaugings. Each domain wall corresponding to
these three minimal gaugings has the same BPS condition. This leads
to the three-fold degeneracy of the 6-form dual of the 2-form
central charge. Given the doublet of 2-form central charges given in
Table \ref{table:centralcharges}, this leads to a total of 6
elementary  half-supersymmetric domain walls.

\bigskip

\noindent {\sl D=7}\,:\ \ In the previous sections we have seen that
there are $20 \subset {\bf {\overline {40}}}$ elementary domain
walls with vector multiplets and $ 5 \subset {\bf {\overline {15}}}$
elementary domain walls with tensor multiplets. From Table
\ref{table:centralcharges} we deduce that the vector  domain walls
have degeneracy 4 while the tensor domain walls have degeneracy 5.
From our general analysis it is easy to see why.

Consider first the vector domain walls. They correspond to minimal
gaugings that are generated by the embedding tensor $\Theta^{MN,P}$.
In subsection 3.3 we found that there are $4\times 5$  minimal
gaugings corresponding to this embedding tensor. The 5 in $4\times
5$ corresponds to the last $SL(5,\mathbb{R})$ index of
$\Theta^{MN,P}$. This direction is similar to the quintet of
membranes that couple to the 3-form potentials $A_{3M}$. In the case
of the membranes this leads to a single orbit of $(n_1,\dots,n_5)$
membranes and a 1-1 relation between membranes and 2-form central
charges. In the case of the domain walls we have an extra direction
corresponding to the first two indices of $\Theta^{MN,P}$. This
extra direction leads to 4 domains walls having the same BPS
condition. This explains the four-fold degeneracy of the vector
domain walls.

We next consider  the tensor domain walls. We found in subsection
3.3 that in this case there are 5 different minimal gaugings leading
to 5 elementary  domain walls which have the same BPS condition.
This leads to the 5-fold degeneracy  of the tensor domain walls.

\bigskip

\noindent {\sl D=6}\,:\ \ One can easily guess how the 5-fold
degeneracy of the 80 elementary domain walls with respect to the 16
central charges arises in six dimensions. The $(D-1)$-forms in six
dimensions belong to the ${\bf 144}$ representation of $SO(5,5)$,
which is the vector-spinor representation. Introducing lightlike
directions $n\pm$, $n=1,...,5$, one can always choose a basis of
Gamma matrices in $SO(5,5)$ such that the highest-weight orbit of
the ${\bf 144}$ is such that for each of the 16 spinor components
only one of the two lightlike directions $n+$ and $n-$ survives, for
any $n$ (see the appendix of \cite{Bergshoeff:2011zk}). This indeed
leads to $16 \times 5 = 80$ elementary domain walls. The 5 in the
$16\times 5$ product is the degeneracy: for each of the 16 spinor
directions, the branes corresponding to the 5 different
non-vanishing lightlike directions lead to the same BPS condition.
\bigskip

One can analyze in a similar way the lower dimensional cases. We leave this as an open project here.

\subsection{BPS conditions and  $E_{11}$}

We can use the knowledge of the roots $\alpha$ corresponding to
supersymmetric domain walls to also determine the Killing spinor
conditions. This rests on the known correspondence between the
variation of the gravitino and the $K(E_{10})$ Dirac
operator~\cite{Damour:2005zs,de Buyl:2005mt,Damour:2006xu}. We
explain this again in the $D=7$ example that was already treated in
the previous section.

{}From an $E_{11}$ perspective the six-forms in the ${\bf
\overline{15}}\oplus {\bf \overline{40}}$ belong to generators that
can be conveniently  described in the $GL(11,\mathbb{R})$
decomposition of $E_{11}$. This decomposition gives an
eleven-dimensional origin to the various six-forms. More precisely,
they come from the following mixed symmetry generators of
$E_{11}$~\cite{Riccioni:2007au}
\begin{align}
\underbrace{E^{a_1\ldots a_6}\phantom{,}}_{\ell=2}, \underbrace{E^{a_1\ldots a_8, b}\phantom{,}}_{\ell=3}, \underbrace{E^{a_1\ldots a_9,b_1b_2b_3}, E^{a_1\ldots a_{10}, b,c}}_{\ell=4}, \underbrace{E^{a_1\ldots a_{10},b_1\ldots b_4, c}\phantom{,}}_{\ell=5}.
\end{align}
Here, we have also indicated the $GL(11,\mathbb{R})$ level~\cite{Damour:2002cu,West:2002jj} in the decomposition. The notation is such that all indices in one block (with the same letter) are antisymmetric. These are irreducible representations so that antisymmetrization including one complete index block and any index from a block to the right gives zero. All these generators give rise to six-forms by putting a sufficient number of indices in the same direction. Putting the maximum number of indices identical will give real roots. This results in the following number of real roots ($\equiv$ number of supersymmetric domain walls) in these generators written in terms of $GL(11,\mathbb{R})$ irreducibles
\begin{subequations}
\label{D=7e11}
\begin{align}
\text{Generator} &&&\text{contains \#($6$-forms)}&&\text{\#(real roots)}\nonumber\\
\label{gen1}
E^{a_1\ldots a_6} &:&& 1&&1,\\
\label{gen2}
E^{a_1\ldots a_8, b} &:&& 24&&12= 4\times 3,\\
\label{gen3}
E^{a_1\ldots a_9,b_1b_2b_3} &:&& 16&&4,\\
\label{gen4}
E^{a_1\ldots a_{10}, b,c} &:&& 10&& 4,\\
\label{gen5}
E^{a_1\ldots a_{10},b_1\ldots b_4, c} &:&& 4&&4=4\times 1.
\end{align}
\end{subequations}
(The Romans mass is contained in the fourth generator (\ref{gen4}),
see
e.g.~\cite{Schnakenburg:2002xx,Kleinschmidt:2003mf,Henneaux:2008nr}.)
The generators in (\ref{gen1}) and (\ref{gen3}) are the ones that
belong to the five supersymmetric (tensor) domain walls in the ${\bf
\overline{15}}$, the remaining ones give $20$ supersymmetric
(vector) domain walls from the ${\bf \overline{40}}$.

We will now use $K(E_{11})$ to determine the supersymmetries that
are preserved by the various branes in terms of their projectors.
For the generators (\ref{D=7e11}) one can compute the action of the
associated $K^*(E_{11})$ generator $J=(E-(E)^\#)/2$ on the
$32$-component supersymmetry parameter in the Dirac
operator~\cite{Damour:2005zs,de
Buyl:2005mt,Damour:2006xu,West:2003fc}. This results in
\begin{subequations}
\begin{align}
\text{Generator} &&&\text{Projector} &&\text{\#(real roots)}\nonumber\\
\label{kgen1}
J^{a_1\ldots a_6} &\rightarrow&& \Gamma^{012345} &&1,\\
\label{kgen2}
J^{a_1\ldots a_8, b} &\rightarrow&& \Gamma^{012345}\Gamma^i &&4\times 3,\\
\label{kgen3}
J^{a_1\ldots a_9,b_1b_2b_3} &\rightarrow&& \Gamma^{012345} &&4,\\
\label{kgen4}
J^{a_1\ldots a_{10}, b,c} &\rightarrow&& \Gamma^{012345} \Gamma^{7\,8\,9\,10}&&4,\\
\label{kgen5}
J^{a_1\ldots a_{10},b_1\ldots b_4, c} &\rightarrow&& \Gamma^{012345}\Gamma^i&&4\times1.
\end{align}
\end{subequations}
Here, the gamma matrices are those of eleven dimensions which are of
size $32\times 32$. The index $i=7,8,9,10$ labels one of the four
compact directions and we aligned the elementary brane along the
directions $0\ldots 5$, leaving $6$ as the transverse direction. We
see that there are different types of projectors. The ones in
(\ref{kgen1}) and (\ref{kgen3}) are the same, leading to a five-fold
degeneracy of the BPS condition for the tensor brane. All five
elementary tensor branes couple to the same central charge.

For the vector branes, there are in total five different projectors.
Four are of the form $\Gamma^{012345}\Gamma^i$ with $i=7,8,9,10$,
and one is of the form $\Gamma^{012345}\Gamma^{7\,8\,9\,10}$. Each
projection condition is four-fold degenerate. In the latter case
(\ref{kgen4}) this is obvious, in the former case one has to combine
three contributions from (\ref{kgen2}) with one from (\ref{kgen5}).
The BPS projectors corresponding to the intersecting brane
configuration presented in eq.~(\ref{14bps}) are given by
(\ref{kgen1}) and (\ref{kgen2}). These projectors are not orthogonal
and combining them yields a projector on a subspace for a $1/4$-BPS
state.

From the form of the root vectors and the $K^*(E_{10})$ Dirac
operator it is also possible to obtain the rescaling of the Killing
spinor that enters the Killing spinor equation. Generally, the
Killing spinor is
\begin{align}
\epsilon = H^{p_\perp/2} \epsilon_0,
\end{align}
where $p_\perp$ is the component of the root (in a metric basis like
in (\ref{d8rt})) along the transverse direction.

A similar analysis can in principle be carried out in all dimensions
in order to determine the supersymmetries preserved by the various
domain walls. The same logic works for other branes than domain
walls as well.

\section{Conclusions}

In this paper we have analyzed the connection between
half-supersymmet\-ric domain walls and deformed maximal supergravity
theories. One of the main results was that  elementary
supersymmetric domain walls exist only in minimally deformed
supergravities. An elementary domain wall was defined as a domain
wall whose dynamics can be described by a supersymmetric worldvolume
action, while a  minimal deformed supergravity was defined as a
supergravity in which the rearrangement of degrees of freedom, which
takes place after the deformation has been turned on, is minimal in
the sense that the minimal number of fields are involved in this
rearrangement. Both the elementary domain walls and the minimal
gaugings  can be characterized in terms of highest-weight orbits
under U-duality. We found that there are many more (non-minimal)
gauged supergravity theories that admit supersymmetric domain-wall
solutions. These  non-elementary domain walls correspond to bound
states of the elementary ones. There exist different types of bound
states. The ones that preserve half supersymmetry, like the
elementary domain walls themselves, are called bound states at
threshold. The ones that preserve less supersymmetry are
non-threshold bound states. The dynamics of these bound states,
threshold and non-threshold, can not be described by a standard
supersymmetric worldvolume action.

We have   elucidated the connection between the elementary domain
walls and the central charges in the supersymmetry algebra. We found
that the relation is many-to-one in contrast to what happens for
branes of codimension three or higher. This extends an earlier
result of ours where we found a two-to-one relation between defect
branes, i.e., branes of codimenion 2, and the central  charges of
the supersymmetry algebra. We have explicitly shown the degeneracy
of the BPS conditions at the level of the domain-wall solutions. The
number of central charges times the degeneracy of the BPS
conditions equals the number of vector minimal gaugings. This
degeneracy, for vector domain walls, is $d+1$ for dimensions $5 \le
D \le 10$, while it is 8 in 4D and 16 in 3D.
The tensor minimal gaugings in 7D, i.e. the ones that
lead to tensor domain-wall solutions, are special in the sense that
they lead to a massive self-dual 3-form. For these domain walls the
degeneracy is 5. Our results could be rephrased in terms of the
$E_{11}$ Kac--Moody algebra.

The fact that there are many domain walls associated to a  given
central charge, and therefore to a given BPS  projection, explains
the fact that there are threshold bound states, preserving the same
amount of supersymmetry as the elementary domain walls. Indeed, a
bound state of two elementary domain walls both satisfying the same
BPS projection does not break supersymmetry any further.
Non-threshold bound states, instead, are bound states of elementary
domain walls satisfying different BPS conditions.

The domain-wall solutions we presented were only local solutions and
do not necessarily have finite energy. Finding rules for a proper
periodic arrangement with orientifolds is an interesting question,
as would be the application of the solutions in the domain
wall/cosmology correspondence mentioned in the introduction.
Furthermore, the application to (A)dS spaces might prove fruitful
since gauged supergravities typically have rather complicated
potentials also allowing for AdS vacua.

Finally, it would  be of interest to extend our analysis to cases
where the original supergravity theory has less than maximal
supersymmetry. We hope to come back to these and other interesting
issues in the future.

\section*{Acknowledgements}

A large part of this work was completed at the Isaac Newton
Institute for Mathematical Sciences (Cambridge). We thank the
organizers of the ``Mathematics and Applications of Branes in String
and M-theory'' programme for their hospitality and financial
support. We wish  to thank Diederik Roest and Henning Samtleben for
useful discussions.


\vskip -20truecm






\end{document}